\documentstyle[psfig]{mn}

\title{The symbiotic binary system RX~Puppis: a possible recurrent nova
with a Mira companion}
\author[J. Miko{\l}ajewska et al.]
   {Joanna~Miko{\l}ajewska$^1$\thanks{e-mail: mikolaj@camk.edu.pl},
    Estela~Brandi$^2$\thanks{Member of the Carrera del Investigator CIC --
    Provincia de Buenos Aires},
    Warren~Hack$^3$, Patricia A.~Whitelock$^4$,
    \newauthor
    Rodolfo Barba$^2$, \thanks{Member of the Carrera del Investigator CONICET}
    Lia Garcia$^2$ and Freddy Marang$^4$\\
   $^1$Copernicus Astronomical Center, Bartycka 18, PL 00-716 Warsaw, Poland\\
   $^2$Facultad de Ciencias Astron{\'o}micas y Geof\'{\i}sicas,
   Universidad Nacional
   de la Plata, Paseo del Bosque S/N, 1900 La Plata, Argentina\\
   $^3$Space Telescope Science Institute, 3700 San Martin Drive, Baltimore,
    MD 21218\\
   $^4$South African Astronomical Observatory, PO Box 9, 7935 Observatory,
   South Africa}

\date{Accepted 1998 ???. Received 1998 ???}
\pagerange{\pageref{firstpage}--\pageref{lastpage}}
\pubyear{1997}

\def\LaTeX{L\kern-.36em\raise.3ex\hbox{a}\kern-.15em
    T\kern-.1667em\lower.7ex\hbox{E}\kern-.125emX}

\begin{document}

\label{firstpage}

\maketitle

\begin{abstract}
 We present an analysis of photometric and spectroscopic observations of the
symbiotic binary system RX~Pup with the aims of developing a reliable binary
model for the system and identifying mechanisms responsible for its
spectacular activity.  The binary is composed of a long-period Mira variable
surrounded by a thick dust shell and a hot $\sim 0.8 \rm M_{\sun}$ white
dwarf companion. The hot component produces practically all activity
observed in the UV, optical and radio range, while variable obscuration of
the Mira by circumstellar dust is responsible for long-term changes in the
near-infrared magnitudes. The observations show RX~Pup underwent a nova-like
eruption during the last three decades. The hot component contracted in
radius at roughly constant luminosity from 1975 to 1986, and was the source
of a strong stellar wind which prevented it from accreting material lost in
the Mira wind. Around 1988/9 the hot component turned over in the HR diagram
and by 1991 its luminosity had faded by a factor of $\sim 30$ with respect
to the maximum plateau value and the hot wind had practically ceased. By
1995 the nova remnant started to accrete material from the Mira wind, as
indicated by a general increase in intensity of the optical continuum and
H\,{\sc i} emission. The quiescent spectrum resembles the quiescent spectra
of symbiotic recurrent novae, and its intensity indicates the hot component
must accrete as much as $\sim 1$ per cent of the Mira wind, which is more or
less the amount predicted by Bondi-Hoyle theory. The earliest observational
records from the 1890s suggest that another nova-like eruption of RX~Pup
occurred around 1894.  \end{abstract}


\section{Introduction}
 Despite numerous papers describing the multi-frequency behaviour of the
symbiotic binary RX~Pup (HD~69190), and several attempts to determine
physical mechanisms responsible for its spectral peregrinations, it remains
one of the most puzzling symbiotic systems ever found.

Based on the peculiar emission-line spectrum, as well as irregular
photometric variability, RX~Pup was initially placed among the irregular
variable stars related to $\eta$~Car or R~CrB (Pickering 1897, 1914;
Payne-Gaposchkin \& Gaposchkin 1938). In 1941 the star was characterised by
a continuous spectrum confined to the yellow and red region and very intense
high excitation emission lines of [Fe\,{\sc VII}], [Ne\,{\sc V}], [Fe\,{\sc
VI}] and [Ca\,{\sc VII}], similar to those observed in CI~Cyg and AX~Per,
which led Swings \& Struve (1941) to classify RX~Pup as a symbiotic star.
The presence of an evolved red giant in RX~Pup was not established
until the 1970s, when M-type absorption features were observed for the
first time (Barton, Phillips \& Allen 1979, Andrillat 1982), while extensive
photometric observations in the near-infrared demonstrated that the system
contains a long period variable with $P \approx 580^{\rm d}$ (Feast,
Robertson \& Catchpole 1977; Whitelock et al. 1983a, hereafter W83). Direct
evidence for the binary nature of RX~Pup has been provided by the detection
of a hot UV continuum source with the {\it International Ultraviolet
Explorer} ({\it IUE}) satellite (Kafatos, Michalitsianos \& Feibelman 1982;
Kafatos, Michalitsianos \& Fahey 1985, hereafter K82 and K85, respectively).

Because of the large volume of observational data available and the
relatively recent onset of activity, RX~Pup has been regarded as a key
object in unravelling the nature of symbiotic Miras; especially as it is one
of the brightest symbiotics at radio and IR wavelengths. The models proposed
to provide the hot component luminosity and to explain the observed activity
involve -- as usual for symbiotic binaries -- either thermonuclear runaway
on the surface of a wind-accreting white dwarf (Allen \& Wright 1988,
hereafter AW88) or variable disk-accretion onto a compact star (Ivison \&
Seaquist 1994, hereafter IS94; see also K85). 

In the AW88 model a low-mass white dwarf slowly accretes hydrogen-rich
material from the wind of a distant Mira companion until a thermonuclear
runaway occurs. Such a flash, especially for a low-mass
white dwarf, can last for many decades; the hot component of RX~Pup now has
a luminosity close to the Eddington value. The observed periods of low and
high excitation were postulated to be caused by variable mass loss from the
white dwarf, while the low reddening reported during the low-excitation
stage was attributed to the close proximity of a shock front (where the hot
component's wind collides with the Mira's wind) to the Mira.

IS94 proposed an alternative scenario in which, during some portion of the
orbital period, the Mira fills its tidal lobe and is gravitationally
stripped of its dusty envelope; as a result the circumstellar reddening is
significantly reduced. An accretion disk or torus forms causing the UV
radiation of the hot companion to be reprocessed. As in the AW88 model, the
bolometric luminosity of the compact star remains unchanged as it goes from
high- to low-excitation phases. In this model the flattening of the radio
spectrum and the disappearance of the high-excitation emission lines result
from cooling of the radiation field in the orbital plane.

Unfortunately, both models have some significant difficulties when
confronted with existing observations. In particular, both AW88 and IS94
assumed that RX~Pup drifts between two extreme -- low and high-excitation --
states without major changes in its bolometric luminosity. In fact there are
major differences between the low-excitation phases reported in 1949/51
(Henize 1976) and the late 1960s, when the optical continuum was very faint
($m_{\rm v} \sim 12.5$, $m_{\rm pg} = 13.5$ in Feb 1968; Sanduleak \&
Stephenson 1973) and the low-excitation phase observed in 1972-1978. The
latter was characterised by a significant increase in the optical brightness
($V \sim 9-8$, $B \sim 9-10$) which strongly suggests intrinsic variability
of the hot source.

This paper contains an analysis of photometric and spectroscopic
observations of RX~Pup made with the objective of developing a reliable
binary model and identifying the causes of its unusual variability.  A
description of the database is given in Section 2, it is analysed in Section
3 and the results discussed in Section 4, while Section 5 contains a
brief summary and conclusion.

\section{Observations}
\subsection{Ultraviolet spectroscopy}

\begin{table}
\caption{Journal of {\it IUE} observations}
\label{}
\footnotesize
\begin{tabular}{ l l c c c}
\hline
~~~Date & JD\,24... & SWP &  LWR/P & $m_{\rm v}$ \cr
\hline
1978, Sep 18 & 43770 & ~2684\,L  & ~2395\,L  \cr
1979, Jun 30 & 44055 & ~5680\,L  & ~4922\,L & ~9.2 \cr
1979, Jul 17 & 44072 & ~5835\,L & ~5081\,L & ~9.3 \cr
1979, Jul 19 & 44074 &          & ~5106\,L & ~9.3 \cr
1980, Apr 15 & 44345 & ~8762\,L & ~7505\,L & ~9.7 \cr
1980, Sep 21 & 44503 & 10189\,L & ~8856\,L & ~9.8\cr
      & & 10190\,L  \cr
      & & 10191\,H \cr
1981, Jun 11 & 44767 & 14239\,L & 10830\,L & 10.2 \cr
      & & 14240\,H & 10832\,L \cr
1982, Mar 22 & 45051 & 16597\,H & 12837\,H & 10.4 \cr
      & & 16598\,L & 12836\,L \cr
1983, Oct 30 & 45638 & 21402\,H & ~2177\,H & 11.0 \cr
      & & 21403\,L & ~2176\,L \cr
1984, Mar 11 & 45771 & 22461\,H & & 11.0 \cr
      & & 22462\,L & ~2924\,L  \cr
1986, May 8 & 46559 & 28270\,H & & 11.2 \cr
      & & 28271\,L & ~8160\,L  \cr
      & &          & ~8161\,L  \cr
      & &          & ~8162\,L \cr
      & &          & ~8163\,L \cr
1986, May 9 & 46560 & 28278\,H & ~8170\,H & 11.2 \cr
      & & 28279\,L &  ~8164\,L  \cr
      & & 28284\,L & ~8165\,L  \cr
      & &          & ~8166\,L \cr
      &           & & ~8167\,L \cr
      &           & & ~8168\,L \cr
      &           & & ~8169\,L \cr
1987, Jul 3 & 46980 & 31285\,H & & 11.3 \cr
1988, Oct 24 & 47459 & 34594\,L & 14309\,L & 11.5 \cr
     &  & 34595\,H & 14310\,H \cr
1989, May 20 & 47667 & 36307\,L & 15552\,L & 11.7 \cr
     &  & 36308\,L \cr
1991, Mar 29 & 48345 & 41209\,L & 20008\,L & 12.2 \cr
1993, May 27 & 49135 & 47757\,L & 25611\,L \cr
1993, May 28 & 49136 &          & 25615\,L \cr
\hline
\end{tabular}
\end{table}

\begin{figure}
\psfig{file=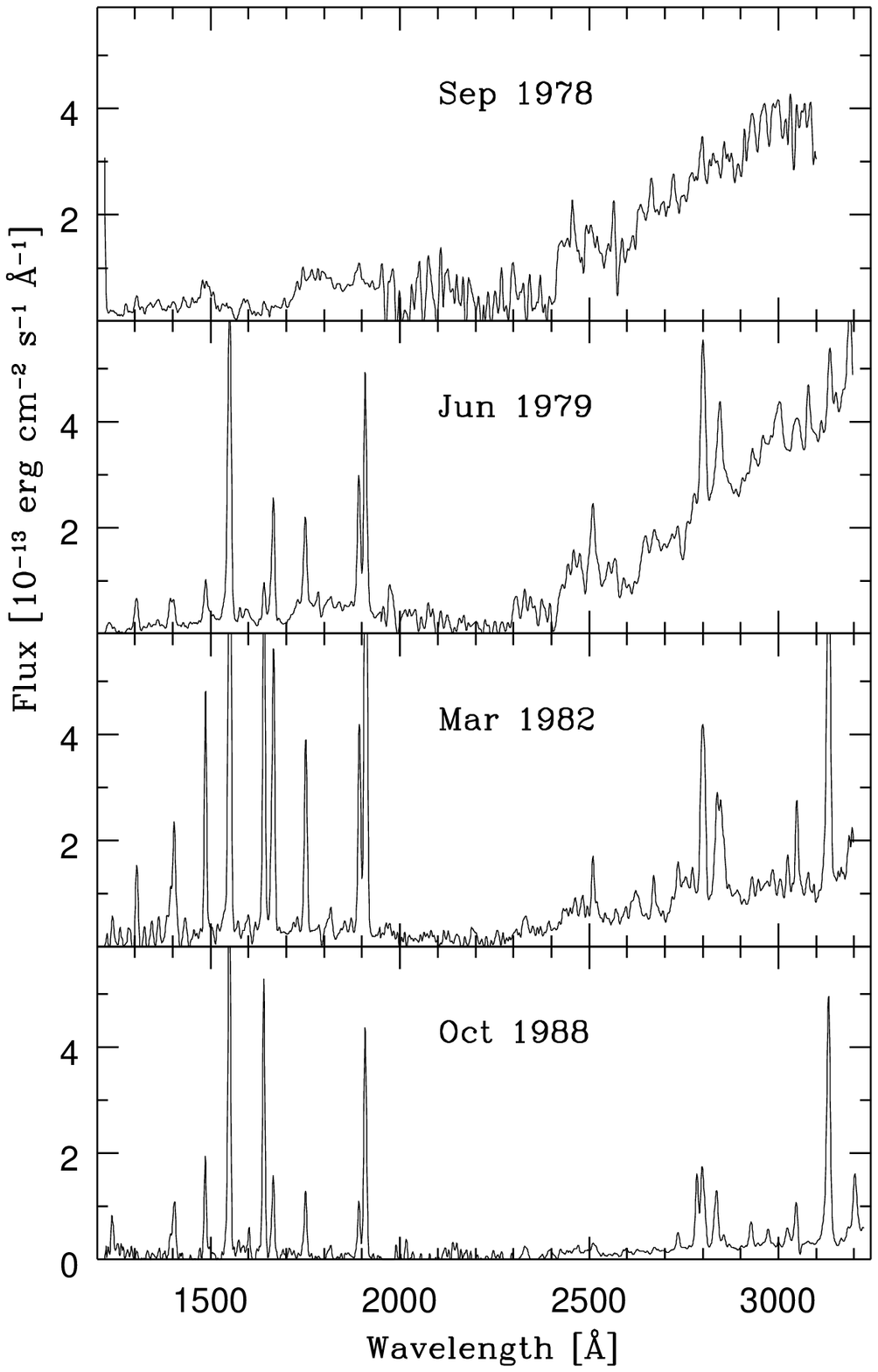,width=0.98\hsize}
\caption{IUE low resolution spectra of RX~Pup taken in 1978-1988.
The spectra show the gradual decline of the UV continuum and the evolution
of emission line fluxes.}
\end{figure}

Various observers acquired ultraviolet spectra of RX~Pup throughout
1978-1993 with {\it IUE}. These spectra covered a significant part of the
last phase of activity. Table 1 summarises the relevant parameters for these
observations, including the Julian Date (JD), the spectrum identifier ({\it L}
and {\it H} indicate low and high resolution data for short wavelength --
SWP -- and long wavelength -- LWR and LWP -- exposures) and visual
magnitudes derived from FES counts. Most of the {\it IUE} data, in
particular the low resolution SWP spectra taken prior to September 1980 and
after May 1986 as well as practically all LWR/LWP spectra, have not been
published.

The low resolution {\it IUE} spectra were transformed into absolute fluxes
using standard calibrations (Holm et al. 1982) and the LWR spectra were
corrected for sensitivity degradation (Clavel, Gilmozzi \& Prieto 1986). The
high resolution spectra were calibrated as suggested by Cassatella, Ponz \&
Selvelli (1982). Table 2 lists emission line fluxes derived by fitting
Gaussian profiles as a function of Julian Date; these estimates have errors
of $\pm 10 \%$ for strong emission lines and $\pm 25 \%$ for weaker lines.
Examples of ultraviolet spectra are shown in Fig. 1.

\begin{table*}
\begin{minipage}{177mm}
\caption{Ultraviolet emission line fluxes in RX~Pup in units
of $10^{-12}\, {\rm erg\,cm^{-2}\,s^{-1}}$}
\label{}
\footnotesize
\begin{tabular}{l r r r r r r r r r r r r r}
\hline
JD 24... & 44055 & 44072 & 44345 & 44503 & 44767 & 45051 & 45638 &
45771 & 46559 & 47459 & 47667 & 48345 \cr
\hline
N\,{\sc v} 1240  & & &  &    & 0.3 & 0.5 & 0.7 & 0.9 & 1.6 & 0.6 & 0.7 \cr
O\,{\sc i} 1304 & 0.8~ & 1.2: & 1.0 & 0.7 & 0.8 & 1.3: & 0.3 &  0.3 &  0.3 &  0.1 \cr
Si\,{\sc iv} 1394 & 0.6:& 0.5:& 0.7 & 0.7 & 0.8 & 0.8 &  0.7 &  0.7 &  0.7 &  0.4 & 0.3 \cr
O\,{\sc iv}] 1403 & 0.5:& 0.8~ & 1.0 & 0.8 & 1.6 & 2.6 & 2.2 & 3.5 & 2.2 &  1.1 &  1.2 \cr
N\,{\sc iv}] 1487 & 0.9~ & 1.4~ & 2.7 & 2.4 & 3.4 & 3.5 & 3.1 & 3.3 &  3.1 &  1.3 &  1.3 \cr
C\,{\sc iv} 1550 & 7.2~ & 8.5~ & 10.7 & 10.6 & 11.7 & 11.3 & 9.5 & 8.7 & 8.5 & 6.5 & 6.2 \cr
[Ne\,{\sc v}] 1575 & & & 0.3 & 0.3 & 0.3 & 0.3 & 0.4 & & 0.3 & 0.3 & 0.2 \cr
[Ne\,{\sc iv}] 1602 & & & 0.6 &    & 0.4 & 0.6 & 0.5 & &  & 0.5 & 0.4 \cr
He\,{\sc ii} 1640 & 0.7~ & 1.1~ &  3.5 &  3.9 &  5.3 &  5.6 & 6.0 & 5.8 & 5.1 & 4.1 & 3.5 \cr
O\,{\sc iii}] 1664 & 2.3~ & 2.9~ &  5.9 &  5.7 &  5.6 &  5.1 & 4.0 & 3.8 & 3.1 & 1.5 & 1.1 \cr
N\,{\sc iv} 1719 & & & & & 0.2 & 0.2 & & 0.3 & 0.2 & 0.1 \cr
N\,{\sc iii}] 1750 & 1.8~ & 2.7~ &  3.0 &  3.6 &  3.7 & 3.4 & 2.7 & 2.5 & 2.0 & 1.2 & 1.1 \cr
Si\,{\sc ii} 1818 & 0.2~ & 0.3~ &  0.5 & 0.5 & 0.4 & 0.5 & 0.5 &   &   & 0.2 & 0.2 \cr
Si\,{\sc iii}] 1892 & 2.6~ & 2.9~ & 3.7 & 3.7 & 3.8 & 3.6 & 2.6 & 2.4 & 2.0 & 1.0 & 0.8 \cr
C\,{\sc iii}] 1909 & 4.1~ & 4.5~ & 8.0 & 8.5 & 10.0 & 10.1 & 8.3 & 8.5 & 5.9 & 3.8 & 3.4 \cr
He\,{\sc ii} 2511, Fe\,{\sc ii} 2513 & 1.1~ & 1.4~ & 1.5 & 0.9 & 1.0 & 1.0 & 0.6 & 0.7 & 0.5 & 0.2 & 0.3 \cr
Al\,{\sc ii}] 2669  &     &     &      & 1.0 & 0.8 & 0.6 & 0.3 & 0.3 & 0.3 & 0.1 & 0.2 \cr
He\,{\sc ii} 2733 &     &  & 0.8 & 0.8 & 0.8 & 0.6 & 0.6 & 0.6 & 0.6 & 0.3 & 0.3 \cr
Fe\,{\sc ii} 2756 &  &  & & 0.6 & & 0.4 & & 0.4 & 0.3 & & 0.1 & 0.1 \cr
Fe\,{\sc ii} 2774  & 0.7~ & 1.3~ & 1.0 & 1.2 & 1.2 & 0.7 & 0.3 & 0.3 \cr
[Mg\,{\sc v}] 2783  &     &     &      &     &     &     & 0.3 & 0.3 & 1.0 & 1.4 & 1.0 \cr
Mg\,{\sc ii} 2800   & 5.3~ & 5.5~ & 3.5 & 3.1 & 8.9 & 9.7 & 7.9 & 4.5 & 3.6 & 2.1 & 1.9 & 0.5 \cr
[Fe\,{\sc iv}] 2829  &     & 0.6: & 0.7 & 0.3 &     & 0.5 & 0.6 & 0.7 & 0.6 & 0.4 & 0.4 \cr
O\,{\sc iii} 2838   &     &      & 0.3 & 1.8 & 2.2 & 1.7 & 1.9 & 1.8 & 1.9 & 1.0 & 0.9 \cr
Fe\,{\sc ii} 2847   & 4.0~ & 3.1~ & 1.9 & 0.9 & 1.5 & 1.6 & 1.0 & 0.7 & 0.4 &  & 0.1 \cr
Fe\,{\sc ii} 2859   &     & 1.2: &     &     & 0.5 & 1.1 & 0.5 & 0.4 & 0.3 &  & 0.3 \cr
[Mg\,{\sc v}] 2928   &     &     &      &     &     &     &  &  & 0.7 & 0.5 & 0.4 \cr
Mg\,{\sc ii} 2936   & 0.6~ & 1.0~ &     & 0.4 & 0.7 & 0.6 \cr
[Ne\,{\sc v}] 2973  &     &     &      &     &     &     & 0.5 & 0.3 & 0.5 & 0.4 & 0.2 \cr
O\,{\sc iii} 3025   &     &     &      & 0.8 & 0.8 & 0.8 & 0.8 & 0.8 & 0.9 & 0.5 & 0.3 \cr
O\,{\sc iii} 3047   &     &     &      & 1.9 & 2.2 & 1.8 & 1.1 & 1.8 & 1.7 & 0.9 & 0.8 \cr
Ti\,{\sc ii} 3079   & 1.4~ & 1.6~ & 1.2 &     & 0.9 & 0.4 &  &  &  &  & 0.3 \cr
O\,{\sc iii} 3133   & 1.9~ & 2.1~ & 6.5 & 5.1 & 8.3 & 9.3 & 10.4& 10.0& 9.1 & 5.4 & 4.7 \cr
\hline
\end{tabular}
\noindent{\footnotesize There were no measurable emission lines present 
on JD\,2\,443\,770 and JD\,2\,449\,135 (see also Section 3.1.2).}
\end{minipage}
\end{table*}

\subsection{HST Observations}

We carried out {\it HST} observations using the FOC/96 camera on 12 March
1991 (JD\,2\,448\,328). 
FOC images were taken with the F501N ([O III] 5007\AA ) filter and the
near-UV objective prism (NUVOP) crossed with the F195W filter. The direct
image taken with the F501N filter provided measurements of the star's
undispersed position for the analysis of the prism observation.

Unfortunately, the FOC images are not conclusive regarding the
point-like nature of RX Pup.  Those images without the prisms in place
are, unfortunately, saturated, leaving only the very extended halo to
hint at the point-like nature of the object.  The best we can say is
that the FOC images place an uncertain upper limit to the size of RX Pup
of approximately $0\farcs 2$.  
Additional UV and optical images taken late in 1993 with
the FOC do not provide substantive evidence of any extended features,
down to an upper limit of $0 \farcs 06$.  However, the low signal-to-noise in
the images and the lack of information regarding the actual focus of
HST and FOC at the time of the observations preclude making any
definitive conclusions.

A relatively long exposure time, of 900 s, was required for the objective
prism image to give sufficiently high signal-to-noise across the full
spectral range from 1750 - 5500 $\rm \AA$. The FOC detectors are count-rate
limited (unlike CCDs) so a $512\times 512$ pixel format was used to maintain
linearity for the star with count rates up to about 0.5 counts/s/pixel. This
format provides a $11\arcsec \times 11\arcsec$ field of view, with pixel sizes of
$0\farcs 022\times 0\farcs 022$.

The objective prism image taken with PRISM2 + F195W (NUVOP) provided low
resolution spectroscopy of RX~Pup (Fig. 2). The peak count rates in the
objective prism image were less than 0.2 counts/s/pixel, insuring
photometric linearity throughout the entire observed spectral region. This
observation gave a signal-to-noise ratio ranging from $\approx$ 2 at 1800
{\rm\AA} to $\approx$ 14 at 4000 {\rm\AA}.

\begin{figure}
\psfig{file=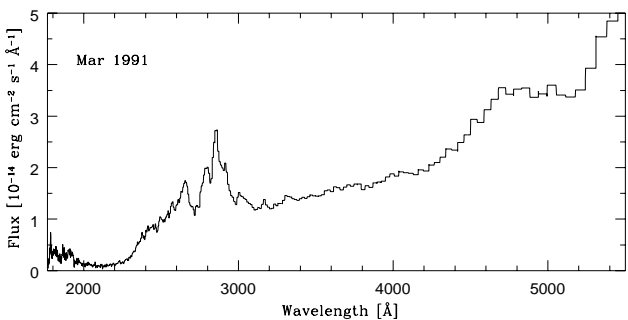,width=0.98\hsize}
\caption{HST low resolution spectrum RX~Pup taken on 12 March 1991.}
\end{figure}

All of the observations were flat-fielded and geometrically corrected using
the standard FOC pipeline calibrations described in detail 
the FOC Instrument Handbook (Nota et al. 1996 henceforth FOCIH).
The dispersed images taken with the objective prisms are aligned
with the visible wavelengths on the lower end and the UV wavelengths at the
upper end.  The NUVOP image had a resolution of 40$\rm \AA$/pixel at
5000$\rm \AA$\ and 0.5$\rm \AA$/pixel at 1700$\rm \AA$ extending across the
entire image (512 pixels).

Seven pixel ($0\farcs 15$) wide regions centred on the peak of the spectrum were
extracted, and the contribution from all the pixels summed to provide the
raw spectrum.  This raw spectrum was wavelength calibrated by applying the
latest dispersion relation for each prism as given in the FOCIH.  
The zero point for the
conversion was based on the position of each star in the F501N image, with
the distance along the spectrum from that zero-point determining the
wavelength of the spectrum. Flux calibrating the spectrum required some
knowledge of the percentage of total flux from the source that was contained
in the spectrum to recover all the flux from the source.  This percentage is
based on the width of the extraction used for the spectrum, which was 7
pixels. An empirical relationship for this percentage was determined using
spectrophotometric standards and the method reported in the FOCIH This
relationship served as the initial estimate for our reductions. With an
estimate of the total flux contained in the extraction, the raw spectrum was
then converted from counts to flux units by dividing out the FOC and filter
sensitivities using software written as part of the Space Telescope Science
Data Analysis System (STSDAS) and described in the FOCIH.

\subsection{Optical spectroscopy}

The optical spectra were taken during 4 observing runs, in November 1990
(JD\,2\,448\,200), April 1991 (JD\,2\,448\,352), June 1995 (JD\,2\,449\,890)
and March 1996 (JD\,2\,450\,167), with the 2.15~m telescope of CASLEO at
San Juan, Argentina. In 1990 and 1991 the Boller \& Chivens Cassegrain
spectrograph and a Reticon photon-counting system known as ``Z-Machine'' was
used. The ``Z-Machine" is a double aperture detector which permits one to
observe simultaneously the sky and the target plus sky. A translation of the
object from one aperture to the other is made at mid exposure for each
observation, and the combination of these two data sets gives a sky-free
spectrum. An aperture of $5''\times 3''$ was used. The 1200 lines mm$^{-1}$
grating gives two wavelength ranges of about $4400-5100 \rm \AA$ and
$5800-7200 \rm \AA$, with dispersions of 0.5 and 1 $\rm \AA$/pixel,
respectively. A spectral resolution of at least 4100 and 2700 in the two
spectral regions, respectively, is indicated by measurements of the FWHM of
arc lines. In 1995 and 1996, the spectra were taken with the REOSC echelle
spectrograph at a resolution of 15\,000, and recorded with a TEK $1024
\times 1024$ CCD. The reduction of the ``Z-Machine" spectra was carried out
with the IHAP image processing software. Wavelength calibration was
performed using the He-Ar-Ne lamps with reference exposures obtained
immediately before and after each stellar exposure, at the same sky
position. Flat-field exposures were also done and a mean value was used in
the reduction procedure. The echelle spectra were reduced with standard IRAF
packages, CCDRED and ONEDSPEC. 

To obtain the flux calibration, standard stars from Stone \& Baldwin (1983)
and Baldwin \& Stone (1984) were observed each night.  A comparison of the
spectra of the standards suggests that the flux calibration errors are about
15 per cent for the ``Z-Machine" and 20 per cent in the central part of each
order for the REOSC echelle images, respectively.

An additional optical spectrum was obtained in May 1991 (JD\,2\,448\,385), 
with the 1.5~m
telescope of ESO at La Silla, and the B\&C spectrograph (used in the range
$4300-6900 {\rm \AA}$) equipped with a CCD ($512 \times 1024$ pixels). The
spectrum was reduced following the IHAP routines, using flatfield exposures,
HeAr wavelength calibration spectra, and standard stars (Hayes \& Latham
1975) observed each night for flux calibrations. The photometric
calibrations are accurate to 10-20 per cent.

Emission line fluxes derived by fitting Gaussian profiles are listed in
Table 3. These data have uncertainties of $\sim$10-15 per cent for strong
lines and $\sim$20-30 per cent for weak lines or noisy spectra. Examples 
are shown in Fig. 3.

\begin{figure}
\psfig{file=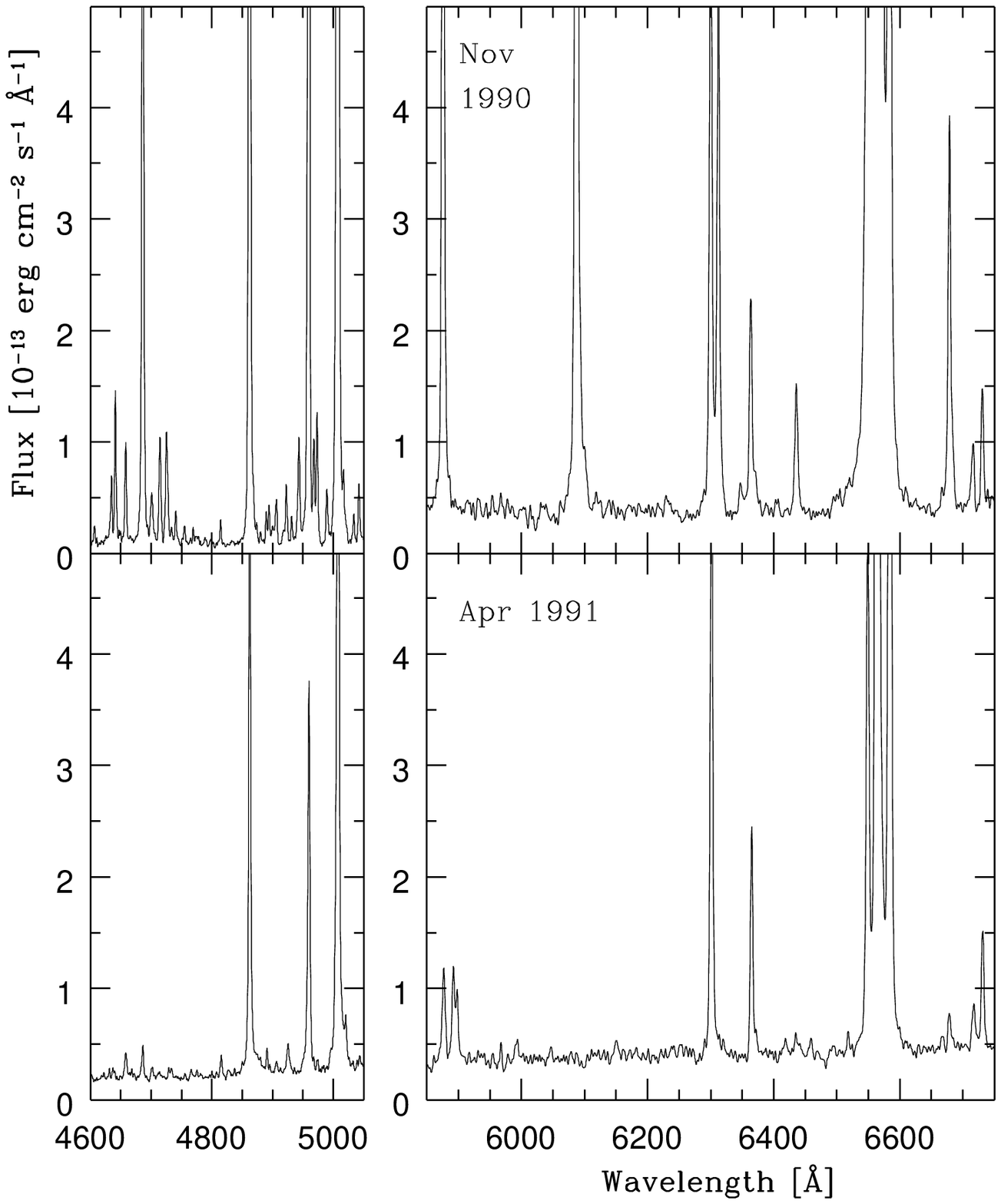,width=0.98\hsize}
\caption{Optical spectra of RX~Pup.}
\end{figure}

\begin{table*}
\begin{minipage}{177mm}
\caption{Optical emission line fluxes in units of $10^{-13}
{\rm erg\,cm\,s}^{-1}$}
\label{}
\footnotesize
\begin{tabular}{l r r r r r l r r r r r}
\hline
JD~24... & 48200 & 48352 & 48385 & 49890 & 50167 &
JD~24... & 48200 & 48352 & 48385 & 49890 & 50167\cr
\hline
[Fe\,{\sc ii}] 4244 &  & & & & 1.0 & 
Fe\,{\sc ii} 5317 & & & 1.5 & 3.3 & 2.7 \cr
C\,{\sc ii} 4267 & & & & & 1.3 &
[Fe\,{\sc ii}], Fe\,{\sc ii} 5362 & & & 0.7 & 2.2 & 1.7 \cr
[Fe\,{\sc ii}] 4287 &  & & & & 1.8 
& [Fe\,{\sc ii}], Fe\,{\sc ii} 5413 & & & 1.1 & 1.2 & 1.5 \cr
$\rm H\gamma$ 4340 &  & & 5.5 & & 6.5 &
Fe\,{\sc ii} 5535 & & & 0.6 & 2.3 & 2.2 \cr
[O\,{\sc iii}] 4363 & & & 2.3 & & 0.7 & 
[O\,{\sc i}] 5577 & & & 0.5 & 0.5: & 1.0: \cr
He\,{\sc i}, Fe\,{\sc ii} 4387 & & & 0.5 & & 0.6 & 
[N\,{\sc ii}] 5755 & & & 8.0 & 1.0 & 0.7 \cr
[Fe\,{\sc ii}] 4414 & 1.6 & 0.8 & 1.4 & & 1.0 & 
He\,{\sc i} 5876 & 61.0 & 4.9 & 4.2 & 0.1: & 0.4 \cr
Fe\,{\sc ii} 4417 & & 1.0 & & & 0.7 &
Na\,{\sc i} 5890 & & 3.7 & 3.1 & abs & abs \cr
[Fe\,{\sc ii}] 4452 &  & 0.4 & 0.7 & & 1.1 &
Na\,{\sc i} 5896 & & 2.6 & 2.8 & abs & abs \cr
[Fe\,{\sc ii}] 4458 &  & 0.5 & 0.6 & & 0.4 &
Fe\,{\sc ii} 5991 & & & 0.6 & 3.0 & 3.9 \cr
He\,{\sc i}, Fe\,{\sc ii} 4472 & 3.1 & 1.1 & 1.1 & & 0.3 &
Fe\,{\sc ii} 6084 & & & 0.4 & 2.3 & 1.9 \cr
[Fe\,{\sc ii}], Fe\,{\sc ii} 4491 &  & 0.6 & 0.8 & & 0.6 
& [Fe\,{\sc vii}], [Ca\,{\sc v}] 6086 & 86.0 & \cr
He\,{\sc ii} 4541 & 1.1 & & & & & Fe\,{\sc ii} 6149 & & & 0.8 & 2.9 & 3.4 \cr
Fe\,{\sc ii} 4584 & & & 1.3 & & 2.8 & Fe\,{\sc ii} 6240 & & & 0.7 & 5.5 & 3.8 \cr
[Fe\,{\sc iii}] 4607 & 0.6 & & & & & Fe\,{\sc ii} 6248 & & & 0.8 & 3.5 & 2.9 \cr
N\,{\sc iii} 4634 & 1.7 & & & & & [O\,{\sc i}] 6300 & 31.8 & 29.0 & 25.1 & 16.9 & 13.0 \cr
N\,{\sc iii} 4641 & 4.0 & & 1.1 & & & [S\,{\sc iii}] 6310 & 29.0 & \cr
[Fe\,{\sc iii}] 4658 & 3.0 & 1.1 & 0.8 & & &
[O\,{\sc i}] 6364 & 11.8 & 10.1 & 8.2 & 3.9 & 4.1 \cr
He\,{\sc ii} 4686 & 39.1 & 1.1 & 1.1 & & &  
Fe\,{\sc ii} 6433 & & & 1.0 & 7.1 & 6.3 \cr
[Fe\,{\sc iii}] 4702 & 2.2 & & 0.2 & & & [Ar\,{\sc v}] 6435 & 7.5 \cr
[Ne\,{\sc iv}], He\,{\sc i} 4714 & 3.8 & & 0.5 & & & 
Fe\,{\sc ii} 6456 & & & 1.2 & 4.4 & 4.8 \cr
[Ne\,{\sc iv}] 4725 & 4.7 & & 0.9 & & & 
Fe\,{\sc ii} 6516 & & & 1.4 & 5.3 & 7.1 \cr
[Fe\,{\sc ii}] 4815 & 0.6 & 0.7 & 0.7 & & 1.1 & 
[N\,{\sc ii}] 6548 &  & 28.1 & 20.6 & 30.0 & \cr
$\rm H\beta$ 4861 & 110.1 & 19.2 & 18.8 & 29.0 & 20.0 & 
$\rm H_{\alpha}$ 6563 & 1790.0 & 173.6 & 145.8 & 700.0 & 630.0 \cr
[Fe\,{\sc ii}] 4890 & 0.8 & 0.5 & 1.0 & 0.8 & 0.6 &
[N\,{\sc ii}] 6584 & 79.1 & 78.2 & 64.4 & 60.0 & 53.4 \cr
[Fe\,{\sc vii}] 4894 & 1.1 & & & & & He\,{\sc i} 6678 & 18.6 & 2.2 & 2.0 & & 1.0 \cr
[Fe\,{\sc ii}] 4905 & 1.7 & & & 0.5 & 0.4 &
[S\,{\sc ii}] 6717 & 3.8 & 2.8 & 2.2 & 1.1 & \cr
He\,{\sc i} 4922 & 2.0 & & & & & [S\,{\sc ii}] 6731 & 5.8 & 5.7 & 4.5 & 2.7 & \cr
Fe\,{\sc ii} 4924 & & 1.5 & 1.6 & 4.0 & 5.5 & 
[Fe\,{\sc iv}], [K\,{\sc iv}] 6793 & 2.1 & \cr
[Fe\,{\sc vii}] 4942 & 3.8 & & & & & He\,{\sc ii} 6891 & 2.4 & \cr
[O\,{\sc iii}] 4959 & 86.9 & 13.0 & 7.2 & & & [Fe\,{\sc iv}] 6997 & 3.6 & \cr
[Fe\,{\sc vi}] 4967 & 3.8 & & & & & [Ar\,{\sc v}] 7006 & 23.2 & \cr
[Fe\,{\sc vi}] 4973 & 4.0 & & & & &  He\,{\sc i} 7065 & 71.1 & 3.9 \cr
[Fe\,{\sc vii}] 4989 & 1.9 & & & & & [Ar\,{\sc iii}] 7136 & 81.8 & & & 0.6 \cr
[Fe\,{\sc ii}] 5006 & & & & 1.1 & 0.9 & 
[Fe\,{\sc ii}] 7155 & 27.5 & & & 5.2 & 2.2 \cr
[O\,{\sc iii}] 5007 & 263.7 & 44.1 & 23.3 & 3.9 & 1.1 & 
[Ar\,{\sc iv}] 7170 & 2.0 &  \cr
He\,{\sc i} 5016 & 1.6 & & & & & [Fe\,{\sc ii}] 7172 & 18.2 & & & 1.3 & 1.9 \cr
Fe\,{\sc ii} 5018 & & & 2.8 & 7.2 & 6.0 & He\,{\sc ii} 7177 & 5.4 \cr
[Fe\,{\sc ii}] 5112 & & & 0.5 & 1.1 & 0.6 & O\,{\sc iii} 7189 & 1.8 \cr
[Fe\,{\sc ii}] 5159 & & & 3.1 & 2.4 & 2.9 & [Fe\,{\sc iv}] 7190 & 3.0 \cr
Fe\,{\sc ii} 5169 & & & 2.3 & 7.8 & 9.0 & [Ca\,{\sc ii}] 7292 & & & & 21.5 & 23.2 \cr
Fe\,{\sc ii}, [N\,{\sc i}] 5198 & & & 1.3 & 2.1 & 2.0 &
[O\,{\sc ii}] 7321 & & & & 3.8 & 5.2 \cr
Fe\,{\sc ii} 5235 & & & 0.9 & 2.2 & 1.9 & [Ca\,{\sc ii}] 7324 & & & & 16.7 & 17.6 \cr
Fe\,{\sc ii} 5285 & & & 0.6 & 2.0 & 2.8 & [O\,{\sc ii}] 7331 & & & & 3.4 & \cr
\hline
\end{tabular}
\end{minipage}
\end{table*}

\subsection{Photometry}

{\it JHKL} broad-band photometry (1.25, 1.65, 2.2, 3.45 $\mu$m) was obtained
with the Mk~II infrared photometer on the 0.75~m telescope at SAAO,
Sutherland (see Carter 1990 for details about the system). Some of the
early data have been published (W83; Whitelock et al. 1984, hereafter W84),
although the values given here (Table 4) might be slightly different from
those published as they have been revised following the work by Carter
(1990). The uncertainty on individual measurements is less than 0.03 mag in
$JHK$, and less than 0.05 mag at $L$. Less precise observations are marked
with a colon.

When available, the FES counts were converted to $V$ magnitudes using the
time dependent correction from Fireman \& Imhoff (1989) and the colour
dependent equations from Imhoff \& Wasatonic (1986). The resulting
magnitudes given in Table 1 have accuracies of 0.1 mag. We have also
collected all published optical photometric data (Eggen 1973; Feast et al.
1977; Klutz 1979; W83, W84; Munari et al. 1992; IS94) as well as visual
magnitude estimates reported on regular basis by the Variable Star Section
of the Royal Astronomical Society of New Zealand (RASNZ). Fig. 4
illustrates the $V/m_{vis}$ light curve together with the IR data.

\begin{figure}
\psfig{file=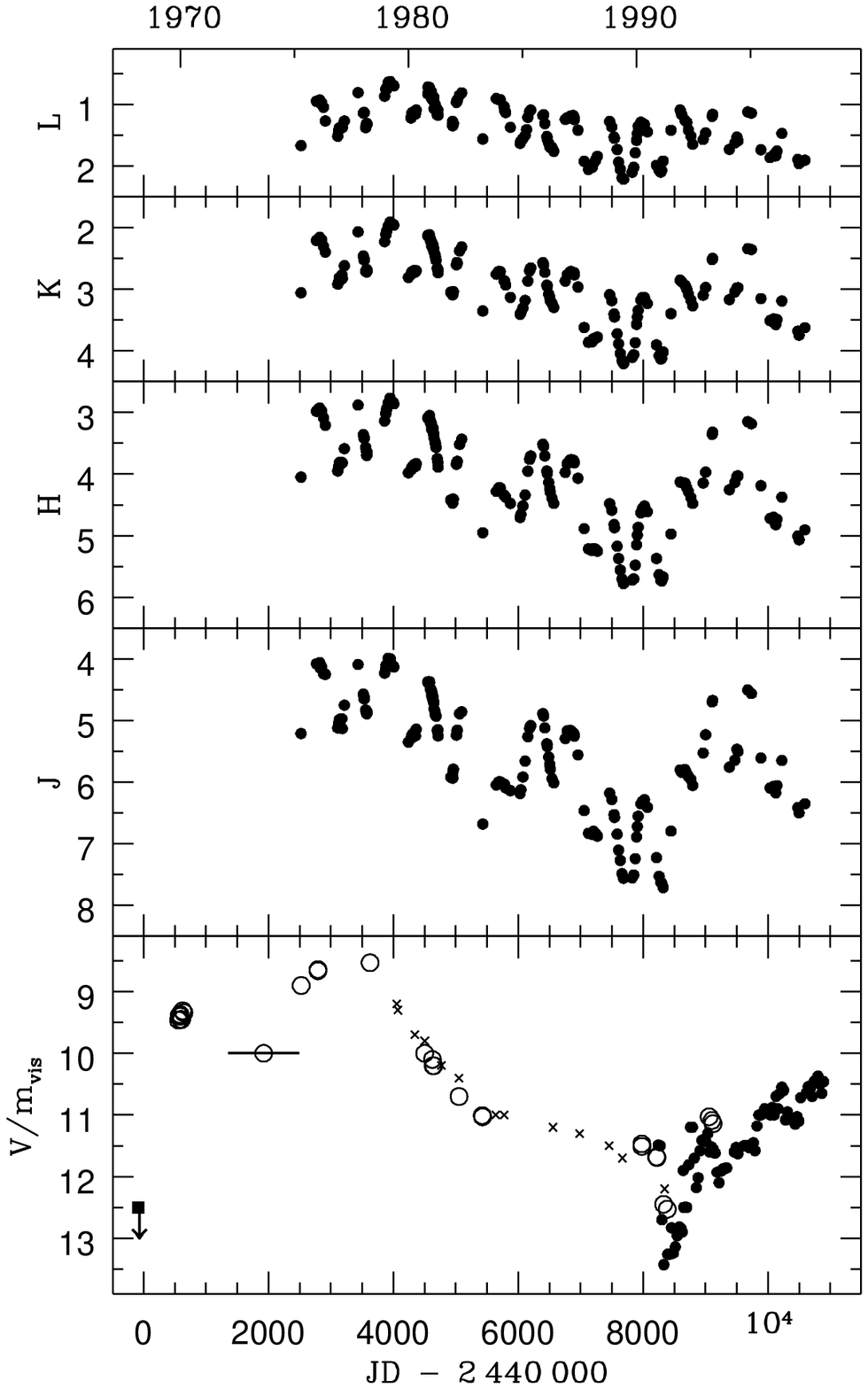,width=0.98\hsize}
\caption{Optical and IR light curves of RX~Pup in 1968-1998. Dots: visual 
observations from RASNZ, open circles: published $V$ mags, crosses: FES mags,
square with a down arrow: February 1968 visual mag limit.}
\end{figure}

\begin{table*}
\begin{minipage}{177mm}
\caption{Infrared photometry of RX~Pup}
\label{}
\footnotesize
\begin{tabular}{l l l l l l l l l l l l l l l}
\hline
 JD~24... & ~~J &  ~~H &  ~~K  & ~~L &  JD~24... & ~~J &  ~~H &  ~~K  & ~~L
&  JD~24... & ~~J &  ~~H &  ~~K  & ~~L \cr
\hline
42520.50 & 5.21  & 4.05  & 3.06  & 1.67 & 44635.36 & 4.62  & 3.26  & 2.29  & 0.92
& 47124.56 & 6.83 & 5.21 & 3.86 & 2.06 \cr
42767.50 & 4.08  & 2.98  & 2.21  & 0.95 & 44639.36 & 4.68  & 3.31  & 2.34  & 0.92
& 47170.48 & 6.83 & 5.24 & 3.86 & 2.02 \cr
42795.50 & 4.09  & 2.95  & 2.19  & 0.95 & 44644.43 & 4.69  & 3.34  & 2.35  & 0.88
& 47188.47 & 6.85 & 5.24 & 3.85 & 2.02 \cr
42818.50 & 4.06  & 2.93  & 2.16  & 0.97  &44647.38 & 4.71  & 3.35  & 2.35  & 0.95
& 47204.41 & 6.80 & 5.21 & 3.80 & 1.96 \cr
42820.50 & 4.08  & 2.96  & 2.19  & 0.93: &44650.38 & 4.81: & 3.41  & 2.40  & 1.07
& 47238.33 & 6.84 & 5.22 & 3.79 & 1.91 \cr
42831.50 & 4.15  & 2.95  & 2.19  & 0.98  &44668.35 & 4.86  & 3.48  & 2.44  & 1.01
& 47267.31 & 6.88 & 5.25 & 3.78 & 1.84 \cr
42851.50 & 4.13  & 2.98  & 2.20  & 0.98  &44671.35 & 4.91  & 3.50  & 2.47  & 1.02
& 47464.61 & 6.18 & 4.48 & 3.09 & 1.28 \cr
42881.50 & 4.24  & 3.09  & 2.31  & 1.05  &44683.32 & 4.93  & 3.57  & 2.53  & 1.06
& 47497.57 & 6.28 & 4.59 & 3.19 & 1.37 \cr
42911.50 & 4.25  & 3.21  & 2.40  & 1.27: &44703.27 & 5.16  & 3.75  & 2.66  & 1.17
& 47533.52 & 6.53 & 4.82 & 3.40 & 1.53 \cr
43109.50 & 5.12  & 3.95  & 2.92  & 1.52  &44712.24 & 5.15  & 3.81  & 2.68  & 1.09
& 47541.41 & 6.57 & 4.87 & 3.45 & 1.55 \cr
43121.50 & 5.04  & 3.88  & 2.89  & 1.43  &44716.23 & 5.25 & 3.89 & 2.73 & 1.18
& 47583.38 & 6.84 & 5.17 & 3.72 & 1.73 \cr
43136.50 & 4.98  & 3.81  & 2.81  & 1.38  &44918.60 & 5.92: & 4.42: & 3.05: &
& 47607.36 & 7.10 & 5.37 & 3.89 & 1.94 \cr
43154.43 & 5.00  & 3.83  & 2.82  & 1.38  &44948.59 & 5.90 & 4.41 & 3.06 & 1.28
& 47634.29 & 7.27 & 5.56 & 4.05 & 2.05 \cr
43174.48 & 4.97  & 3.81  & 2.77  & 1.37  &44949.52 & 5.91 & 4.46 & 3.09 & 1.34
& 47661.26 & 7.49 & 5.70 & 4.16 & 2.19 \cr
43182.39 & 5.13  & 3.82  & 2.83  & 1.38  &44950.56 & 5.93 & 4.47 & 3.09 & 1.35
& 47687.21 & 7.56 & 5.78 & 4.21 & 2.22 \cr
43215.32 & 4.75  & 3.59  & 2.62  & 1.27  &44955.55 & 5.82 & 4.41 & 3.05 & 1.31
& 47823.61 & 7.56 & 5.72 & 4.11 & 2.10 \cr
43434.60 & 4.09  & 2.88  & 2.07  & 0.81  &44960.50 & 5.79 & 4.41 & 3.04 & 1.31
& 47850.53 & 7.51 & 5.70 & 4.06 & 2.02 \cr
43519.50 & 4.57  & 3.36  & 2.46  & 1.14  &45010.46 & 5.24 & 3.84 & 2.61 & 0.97
& 47873.53 & 7.24 & 5.48 & 3.87 & 1.79 \cr
43530.50 & 4.61  & 3.40  & 2.50  & 1.14  &45017.40 & 5.18 & 3.80 & 2.57 & 0.94
& 47894.54 & 6.89 & 5.15 & 3.57 & 1.59 \cr
43532.48 & 4.65  & 3.42  & 2.53  & 1.13  &45022.38 & 5.16 & 3.79 & 2.58 & 0.94
& 47907.47 & 6.72 & 4.99 & 3.45 & 1.48 \cr
43556.47 & 4.83  & 3.57  & 2.71  & 1.38  &45060.29 & 4.89 & 3.52 & 2.38 & 0.85
& 47920.46 & 6.55 & 4.86 & 3.34 & 1.36 \cr
43572.   & 4.88  & 3.70  & 2.72  & 1.33  &45095.20 & 4.86 & 3.43 & 2.32 & 0.81
& 47964.35 & 6.35 & 4.63 & 3.18 & 1.29 \cr
43572.34 & 4.85  & 3.64  & 2.68  & 1.33  &45432.25 & 6.68 & 4.95 & 3.35 & 1.56
& 47989.28 & 6.31 & 4.56 & 3.13 & 1.31 \cr
43573.   & 4.89  & 3.69  & 2.71  & 1.31  &45648.60 & 6.05 & 4.28 & 2.76 & 0.91
& 48022.33 & 6.29 & 4.52 & 3.14 & 1.33 \cr
43573.35 & 4.87  & 3.64  & 2.69  & 1.32  &45687.46 & 5.99 & 4.22 & 2.71 & 0.93
& 48068.19 & 6.41 & 4.61 & 3.23 & 1.45 \cr
43860.56 & 4.23  & 3.14  & 2.23  & 0.87  &45713.34 & 6.00 & 4.23 & 2.72 & 0.93
& 48213.52 & 7.22 & 5.37 & 3.90 & 1.99 \cr
43878.54 & 4.15  & 3.02  & 2.11  & 0.75  &45773.   & 6.06 & 4.34 & 2.86 & 1.03
& 48256.56 & 7.53 & 5.63 & 4.08 & 2.07 \cr
43881.47 & 4.10  & 3.00  & 2.11  & 0.76  &45778.   & 6.03 & 4.35 & 2.87 & 1.10
& 48281.44 & 7.62 & 5.72 & 4.13 & 2.10 \cr
43890.52 & 4.10  & 2.95  & 2.06  & 0.76  &45800.   & 6.09 & 4.38 & 2.93 & 1.14
& 48296.41 & 7.64 & 5.74 & 4.12 & 2.07 \cr
43896.46 & 4.09  & 2.93  & 2.05  & 0.73  &45874.   & 6.14 & 4.48 & 3.13 & 1.37
& 48321.40 & 7.71 & 5.67 & 4.02 & 1.92 \cr
43918.44 & 3.99  & 2.84  & 1.97  & 0.64  &46029.50 & 6.19 & 4.70 & 3.41 & 1.63
& 48446.19 & 6.79 & 4.97 & 3.40 & 1.42 \cr
43922.43 & 4.03  & 2.85  & 1.97  & 0.68  &46042.55 & 6.13 & 4.65 & 3.38 & 1.60
& 48593.60 & 5.80 & 4.13 & 2.85 & 1.09 \cr
43947.32 & 4.00  & 2.77  & 1.92  & 0.67  &46074.50 & 5.92 & 4.52 & 3.30 & 1.56
& 48616.53 & 5.84 & 4.14 & 2.88 & 1.17 \cr
43949.33 & 4.00  & 2.78  & 1.91  & 0.63  &46110.37 & 5.66 & 4.34 & 3.18 & 1.50
& 48670.48 & 5.80 & 4.15 & 2.94 & 1.26 \cr
43952.31 & 4.04  & 2.82  & 1.95  & 0.66  &46133.39 &   &   &   & 1.41
& 48700.37 & 5.86 & 4.23 & 3.00 & 1.29 \cr
44001.23 & 4.12  & 2.84  & 1.95  & 0.70  &46153.32 & 5.26 & 3.95 & 2.87 & 1.21
& 48725.24 & 5.91 & 4.29 & 3.09 & 1.42 \cr
44008.23 & 4.13  & 2.86  & 1.96  & 0.70  &46181.27 & 5.13 & 3.76 & 2.70 & 1.12
& 48765.22 & 5.96 & 4.39 & 3.18 & 1.52 \cr
44240.51 & 5.35  & 3.98  & 2.81  &   & 46199.24 & 5.08 & 3.71 & 2.66 & 1.09
& 48792.19 & 6.05 & 4.48 & 3.27 & 1.65 \cr
44282.40 & 5.29  & 3.93  & 2.76  & 1.22  &46393.59 & 4.89 & 3.52 & 2.57 & 1.18
& 48961.52 & 5.53 & 4.15 & 3.10 & 1.57 \cr
44295.37 & 5.24  & 3.89  & 2.72  & 1.14  &46407.60 & 4.93 & 3.55 & 2.61 & 1.17
& 49001.34 & 5.23 & 3.97 & 2.97 & 1.46 \cr
44308.37 & 5.22  & 3.88  & 2.72  & 1.14  &46425.60 & 5.12 & 3.71 & 2.73 & 1.31
& 49106.23 & 4.70 & 3.36 & 2.52 & 1.19 \cr
44317.36 & 5.23  & 3.87  & 2.71  & 1.16  &46458.46 & 5.38 & 3.95 & 2.94 & 1.52
& 49113.20 & 4.67 & 3.32 & 2.50 & 1.16 \cr
44327.29 & 5.21  & 3.87  & 2.70  & 1.16  &46465.44 & 5.42 & 3.99 & 2.96 & 1.57
& 49381.47 & 5.76 & 4.26 & 3.17 & 1.73 \cr
44331.26 & 5.20  & 3.85  & 2.71  & 1.12  &46488.46 & 5.59 & 4.14 & 3.08 & 1.63
& 49468.34 & 5.64 & 4.14 & 3.04 & 1.63 \cr
44353.24 & 5.25  & 3.88  & 2.72  & 1.16  &46503.33 & 5.71 & 4.23 & 3.14 & 1.68
& 49503.25 & 5.47 & 4.04 & 2.98 & 1.53 \cr
44364.23 & 5.14  & 3.83  & 2.70  & 1.09  &46514.32 & 5.79 & 4.30 & 3.19 & 1.70
& 49519.19 & 5.50 & 4.03 & 2.98 & 1.59 \cr
44551.61 & 4.38  & 3.09  & 2.13  & 0.83: &46540.35 & 5.94 & 4.40 & 3.23 & 1.70
& 49676.53 & 4.50 & 3.15 & 2.34 & 1.12 \cr
44554.57 & 4.37  & 3.09  & 2.13  & 0.72  &46569.21 & 6.01 & 4.48 & 3.30 & 1.76
& 49734.56 & 4.56 & 3.19 & 2.36 & 1.14 \cr
44578.58 & 4.37  & 3.05  & 2.12  & 0.73  &46752.59 & 5.29 & 3.97 & 2.87 & 1.25
& 49887.22 & 5.61 & 4.19 & 3.15 & 1.74 \cr
44582.56 & 4.38  & 3.10  & 2.15  & 0.75  &46783.50 & 5.17 & 3.83 & 2.77 & 1.22
& 50034.55 & 6.10 & 4.72 & 3.51 & 1.87 \cr
44593.52 & 4.48  & 3.17  & 2.21  & 0.78  &46825.42 & 5.16 & 3.78 & 2.73 & 1.21
& 50089.50 & 6.07 & 4.70 & 3.48 & 1.83 \cr
44596.55 & 4.50  & 3.16  & 2.20  & 0.81  &46833.43 & 5.19 & 3.79 & 2.75 & 1.22
& 50111.43 & 6.13 & 4.78 & 3.53 & 1.81 \cr
44606.54 & 4.51  & 3.18  & 2.23  & 0.79  &46844.44 & 5.16 & 3.77 & 2.71 & 1.19
& 50123.49 & 6.18 & 4.82 & 3.57 & 1.84 \cr
44613.50 & 4.59  & 3.28  & 2.30  & 0.91  &46881.33 & 5.20 & 3.77 & 2.73 & 1.18
& 50142.46 & 6.06 & 4.74 & 3.49 & 1.76 \cr
44614.54 & 4.54  & 3.23  & 2.25  & 0.87  &46897.29 & 5.23 & 3.81 & 2.74 & 1.22
& 50221.28 & 5.65 & 4.38 & 3.19 & 1.47 \cr
44620.49 & 4.60  & 3.24  & 2.28  & 0.88  &46897.33 & 5.25 & 3.82 & 2.77 & 1.25
& 50478.48 & 6.41 & 5.01 & 3.68 & 1.90 \cr
44627.41 & 4.65  & 3.26  & 2.33  & 0.89  &46955.26 & 5.56 & 4.07 & 2.97 & 1.42
& 50498.45 & 6.50 & 5.07 & 3.75 & 1.96 \cr
44631.42 & 4.60  & 3.27  & 2.30  & 0.92  &47055.65 & 6.46 & 4.89 & 3.62 & 1.93
& 50590.23 & 6.35 & 4.91 & 3.63 & 1.91 \cr
\hline
\end{tabular}
\end{minipage}
\end{table*}

\section{Analysis}

\subsection{Variability}

\subsubsection{A brief survey of observations prior to 1969}

Although the first observational records of RX~Pup date back to the end of
XIX century, the star has been generally neglected by most observers, and
only sketchy information on its spectroscopic and photometric variations
prior to the early 1970s is available. 

RX~Pup was classed among ``stars having peculiar spectra" by Fleming (1912).
The spectrum taken on November 24, 1894 resembled that of $\eta$~Car,
showing bright H$\beta$, H$\gamma$, H$\delta$ and H$\zeta$ (Pickering 1897).
Another spectrum taken on January 8, 1897, showed in addition to the strong
H\,{\sc} Balmer lines, bright He\,{\sc ii} $\lambda$4686, [O\,{\sc iii}]
$\lambda$4363 and [Ne\,{\sc iii}] $\lambda$3869 (Cannon 1916). Cannon also
noted that the spectral lines (although not their widths) resembled those of
novae in the nebular phase. Extensive photometry obtained in 1890-1924
(Pickering 1914; Yamamoto 1924) showed the star brightened from $m_{\rm pg}$
= 12 in 1890 to $m_{\rm pg}$ = 11.13 in 1894, then declined to a very deep
minimum ($m_{\rm pg}$ = 14.1) in 1904-5. After recovery c.1909, the
photographic magnitude showed only small fluctuations between 11.6 and 12.3
mag. 

The next spectra of RX~Pup, obtained in January 1941, revealed a continuum
confined to the yellow and red region, and very strong high excitation
emission lines of [Fe\,{\sc vii}], [Ne\,{\sc v}], [Fe\,{\sc vi}] and
[Ca\,{\sc vii}], in addition to strong H\,{\sc i} Balmer series, He\,{\sc
ii}, [Ne\,{\sc iii}], and [O\,{\sc iii}] emission lines (Swings \& Struve
1941). Swings \& Struve also pointed out the spectroscopic similarity of
RX~Pup to the symbiotic stars CI~Cyg and AX~Per, although they did not find
definite evidence for a late-type component in RX~Pup.

The spectra taken between 1949 and 1951 showed very faint (or absent)
continuum, moderately sharp H$\alpha$, as well as [O\,{\sc i}]
$\lambda\lambda$6300, 6364 and other blended emissions (most of them
probably Fe\,{\sc ii} lines) between $\lambda\,6300\, \rm \AA$ and H$\alpha$
(Henize 1976). A similar low ionisation spectrum, with no He\,{\sc ii}
$\lambda$4686, [O\,{\sc iii}] $\lambda$5007 very weak relative to H$\beta$
and a very faint continuum, was observed on February 26, 1968 by Sanduleak
\& Stevenson (1973) who also estimated the visual and photographic
magnitudes at $m_{\rm vis} \ga 12.5$ and $m_{\rm pg} = 13.5 \pm 0.5$,
respectively.

\subsubsection{The spectral transformations in 1969-1996}

Between 1968 and 1969 the optical brightness of RX~Pup increased by a few
magnitudes (Fig. 4). Between December 1969 and February 1970, the star
reached $V \sim 9.5 - 9.3$ (Eggen 1973).  Its colours $B-V \sim 1.3 - 1.2$,
$U-B \sim 0.5 - 0.4$, and $R-I \sim 0.8$, corrected for a reasonable amount
of the interstellar reddening, E$_{B-V} \sim 0.6$ (Section 3.2), indicated
the presence of a bright blue continuum. Optical spectroscopy obtained in
February 1972 and March 1975 (Swings \& Klutz 1976; also Webster \& Allen
1975) revealed a bright late A- or early F-type continuum with strong
H\,{\sc i} as well as O\,{\sc i}, [O\,{\sc i}], Ca\,{\sc ii}, [Ca\,{\sc
ii}], and numerous Fe\,{\sc ii} and [Fe\,{\sc ii}] emission lines. RX~Pup
maintained this low excitation emission spectrum together with a strong blue
continuum until March 1979 (Klutz, Simonetto \& Swings 1978; Klutz \& Swings
1981). The optical brightness reached a maximum in 1977-78, with $V \sim
8.5$, $B-V \sim 0.9$, and $U-B \sim 0.1-0.5$, the optical spectrum resembled
that of a Be or shell star (Feast et al. 1977; Klutz 1979). The first {\it
IUE} spectrum taken in September 1978 (Fig. 1) shows a UV continuum like
that of a Be/shell star. Similar optical/UV spectra have been observed in
CH~Cyg and other symbiotic stars during the visual maximum of their
outbursts (Miko{\l}ajewska, Selvelli \& Hack 1988; Kenyon 1986).

The H\,{\sc i} Balmer lines presented double P Cygni profiles (V $<<$ R) in
1972, which evolved into a complex P~Cygni structure in 1975, with one
emission and a series of absorption components ranging from --1100 to --230
$\rm km\,s^{-1}$ (Swings \& Klutz 1976). Such complex structure was also
observed in 1976, although higher members (such as H$\gamma$ and H$\delta$)
showed very little absorption. In 1978 the Balmer lines became fairly stable
emission lines with a sharp P Cygni profiles and a few fainter absorptions
with positive velocities (Klutz 1979).

In December 1979 RX~Pup returned to the high excitation conditions exhibited
40 years earlier; the appearance of strong and broad emission lines of
He\,{\sc ii}, N\,{\sc iii} and [O\,{\sc iii}] in the optical range was
accompanied by decline in the $UBV$ fluxes (Klutz \& Swings 1981; see
Fig. 4). The Balmer continuum strengthened, and the intensities of the
$\lambda$4640 feature and He\,{\sc ii} $\lambda$4686 emission resembled
those observed in WN7-8 stars. The 1979 spectrum, however, showed neither
[Fe\,{\sc vi}] nor [Fe\,{\sc vii}]. The next optical spectrum taken in April
1983 revealed faint [Fe\,{\sc vi}] and [Fe\,{\sc vii}] emission lines (W83),
with a flux ratio [Fe\,{\sc vii}]\,$\lambda$6086/He\,{\sc i}\,$\lambda$5876
$\sim 0.2$. The intensity of the [Fe\,{\sc vii}] was growing, and in the
1984/89/90 spectra (AW88; de Freitas Pacheco \& Costa 1992, hereafter FPC92;
Table 3) the [Fe\,{\sc vii}]\,$\lambda$6086/He\,{\sc i}\,$\lambda$5876 flux
ratio increased to 0.3/1.7/1.4, respectively. The changes in the optical
spectrum observed between 1979 and 1989 clearly indicate a gradually
increasing degree of ionisation, similar to that reported for symbiotic
novae (e.g. Kenyon 1986).

\begin{figure}
\psfig{file=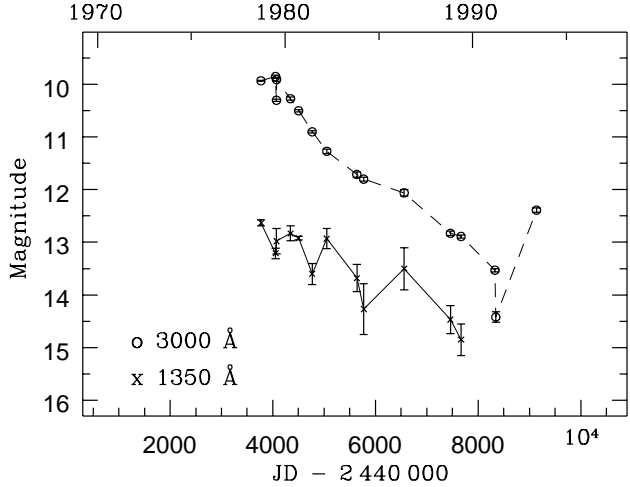,width=0.98\hsize}
\caption{Variation of UV continuum magnitudes in 1978-1993.}
\end{figure}
\begin{figure}
\psfig{file=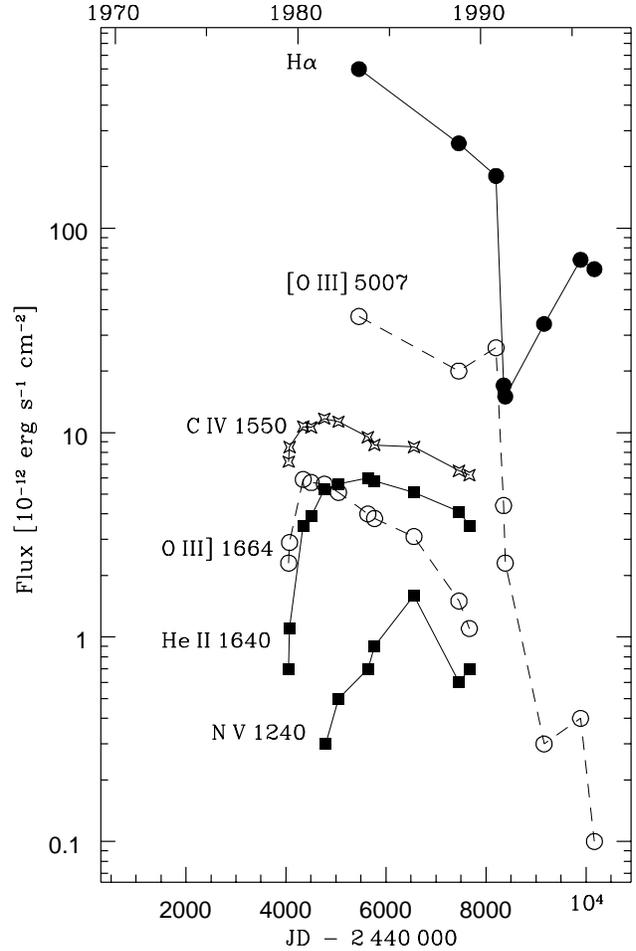,width=0.98\hsize}
\caption{Evolution of optical and UV emission line fluxes in 1979-1996.
The H$\alpha$ and [O\,{\sc iii}]\,$\lambda$5007 fluxes published by W83,
van Winckel, Duerbeck \& Schwarz (1993) and IS94 are also plotted.}
\end{figure}

The evolution of the blue continuum and emission line spectrum in 1978-1993
can be traced best using {\it IUE} data. Fig. 1 illustrates the changes in
the UV. To establish the continuum behaviour, we measured average fluxes,
$F_{\lambda}$, in $20\, \rm \AA$ bins free from emission lines and derived
magnitudes using: $m_{\lambda} = -2.5 \log F_{\lambda} - 21.1$. The
continuum declines steadily from 1978 to 1991, in a similar way to the
visual light (Figs 5 \& 4); the UV continuum seems to fade less at short
wavelengths ($\lambda\lambda \la 1500 \rm \AA$) than at longer wavelengths
($\lambda\lambda \la 2500 \rm \AA$). The continuum decline is matched by an
increase in flux from most of the emission lines (Fig. 6). In particular,
the He\,{\sc ii} $\lambda$1640 flux increased by a factor of $\sim 5$
between June 1979 and April 1980. The He\,{\sc ii} flux continued a slower
increase until June 1981 and from then to the observation of May 1986 it
remained at roughly constant level. Similar behaviour was seen amongst other
medium excitation lines, such as C\,{\sc iv} $\lambda$1550, N\,{\sc iv}]
$\lambda$1487, Si\,{\sc iv} $\lambda$1394 and O\,{\sc iv}] $\lambda$1403.
N\,{\sc v} $\lambda$1240 appeared for first time in June 1981, while
[Mg\,{\sc v}] $\lambda$2783 appeared in October 1983 (Fig. 7); these lines
showed maximum fluxes in the spectra taken during 1986 and 1988,
respectively. Generally, the trends observed in the UV continuum and
emission line fluxes show an evolution towards higher stages of ionisation,
similar to that observed in symbiotic novae (Nussbaumer \& Vogel 1990;
M{\"u}rset \& Nussbaumer 1994).

\begin{figure}
\psfig{file=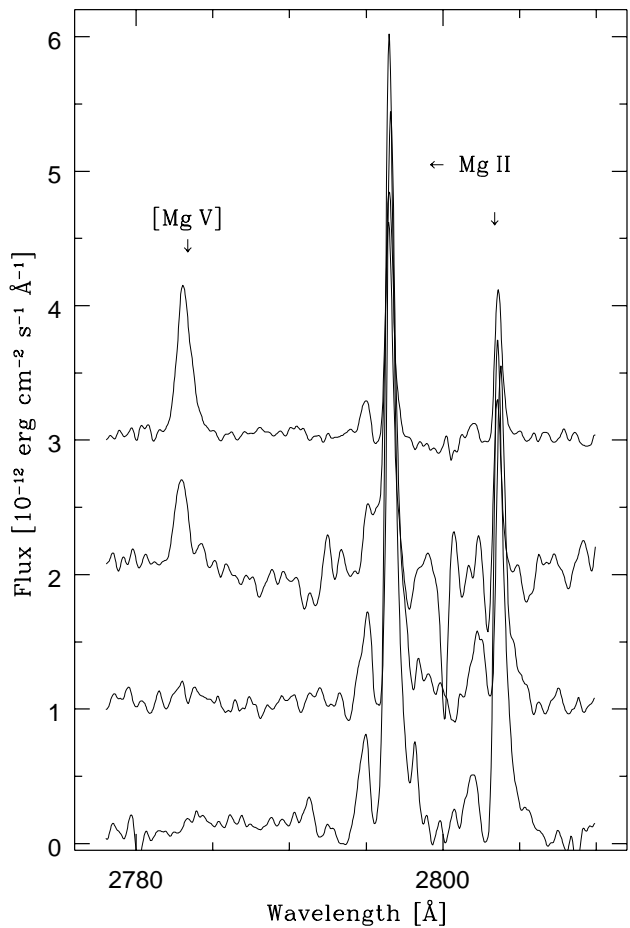,width=0.98\hsize}
\caption{Appearance and evolution of [Mg\,{\sc v}].
The high resolution spectra were taken on (from bottom to top):
22 March 1982, 30 October 1983, 9 May 1986 and 25 October 1988,
respectively, show the appearance and increase of [Mg\,{\sc v}] $\lambda$2783.}
\end{figure}

From an analysis of high resolution {\it IUE} spectra taken between
September 1980 and March 1984, K85 (also K82) found that many of the strong
emission lines exhibit multiple component structure. Our analysis, however,
does not confirm some of their identifications. For example, the redshifted
component of He\,{\sc ii} $\lambda$1640 as well as the blue-shifted
component of N\,{\sc iii}] $\lambda$1749.7 (Fig. 1b and Fig. 1d of K85) are
misidentified O\,{\sc i}] $\lambda$1641.4 and N\,{\sc iii}] $\lambda$1748.6, 
respectively. The intensity of the $\lambda$1641 feature is
strongly correlated with the O\,{\sc i} $\lambda$1304 line flux measured on
low resolution spectra; this feature is absent on all high resolution
spectra taken in 1983-1988 when the O\,{\sc i} $\lambda$1304 was very faint
or absent. Moreover, profiles of all semiforbidden emission lines detected
on the high resolution spectra show more or less the same velocity
structure, with the width at the foot of $\sim 250\, \rm km\,s^{-1}$
centred roughly at the stellar velocity ($\sim 10\, \rm km\,s^{-1}$;
Wallerstein 1986). The only exception would be N\,{\sc iii}] $\lambda$1749.7
if the K85 interpretation were accepted. We also disagree with K85 about the
close similarity between the C\,{\sc iv} and He\,{\sc ii} profiles. The
C\,{\sc iv} $\lambda\lambda$1548.2, 1550.8 doublet emission profile is
truncated by a narrow P Cygni structure at $\sim -90\, \rm km\,s^{-1}$, and
it exhibits rather complex structure, while He\,{\sc ii} has a considerably
simpler profile which is almost symmetric around the stellar velocity.

The C\,{\sc iv} doublet shows unusual intensities, the
$I(\lambda1548.2)/I(\lambda1550.8)$ ratio had an observed value of $\sim
0.6$ in 1980-82, then in 1983 its value increased to $\sim 0.8$. In 1987 the
C\,{\sc iv} doublet ratio reached the optical thick limit of unity (see also
Michalitsianos et al. 1988), while in 1988
$I(\lambda1548.2)/I(\lambda1550.8)\simeq 1.1$. The C\,{\sc iv} ratio seems
to increase following the general enhancement of high-excitation emission
line fluxes (Table 2; Fig. 6). Similar behaviour was reported in the
symbiotic nova HM Sge (M{\"u}ller \& Nussbaumer 1985). Michalitsianos, Kafatos
\& Meier (1992) demonstrated that the combined fluxes of Fe\,{\sc ii} 
Bowen-pumped lines can account for an appreciable portion of the flux deficit 
in the C\,{\sc iv} $\lambda$1548.2 line in 1980-83, when the C\,{\sc iv} 
doublet ratio is less than the optically thick limit. Their interpretation is
consistent with the presence of O\,{\sc i}] $\lambda$1641.3 in 1980-83
indicating the emitting region was optically thick.

By early 1991, the high excitation emission lines had practically
disappeared. The HST spectrum taken in March 1991 (Fig. 3) revealed a very 
faint UV continuum without strong emission lines. In particular, the
intercombination Si\,{\sc iii}]\,$\lambda$1892 and C\,{\sc
iii}]\,$\lambda$1909 still strong on the SWP spectrum of May 1989 had
disappeared, as had the O\,{\sc iii}\,$\lambda$3133 emission which was very
strong in the 1980s.  The only emission detected in the UV range was a weak
Mg\,{\sc ii} $\lambda$2800. The optical continuum was also faint, consistent
with the deep minimum shown by the $V/m_{\rm vis}$ light curve at this time
(Fig. 4). The lack of measurable short-wavelength continuum or emission
lines (except Mg\,{\sc ii}\,$\lambda$2800 and very faint Fe\,{\sc
ii}\,$\lambda$2756) was confirmed by the {\it IUE} spectra taken a few days
after the HST data. The optical spectra taken in April and May 1991, also
showed a significant decline in the continuum and emission line fluxes (Fig. 3
and Fig. 6). The only emission lines that remained practically unchanged
were low excitation forbidden lines [O\,{\sc i}], [N\,{\sc ii}] and
[Fe\,{\sc ii}]. The 1991 spectrum was also characterised by the appearance
of Na\,{\sc i}\,D  in emission (Fig. 8). The decrease in the
He\,{\sc ii}/H$\beta$ ratio from $\sim 0.4$ in the 1990s to $\sim 0.06$ in
1991 suggests a general decline of the source of ionisation.

\begin{figure}
\psfig{file=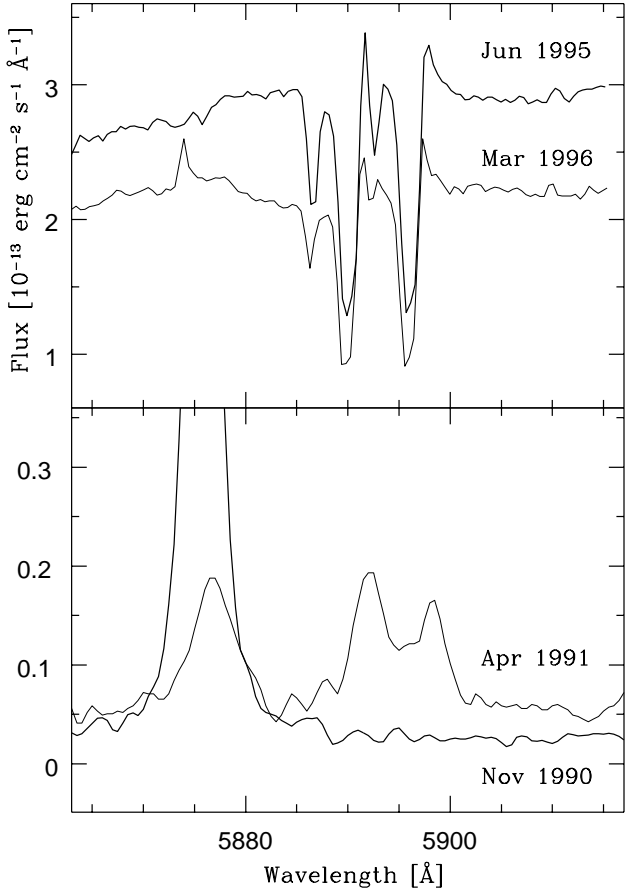,width=0.98\hsize}
\caption{Evolution of Na\,{\sc i} D$_1$,\,D$_2$ lines.}
\end{figure}

Between 1992 and 1993, the optical magnitudes recovered to about the same
level as in the late 1980s (Fig. 4), the degree of excitation, however,
continued to decrease. He\,{\sc ii}\,$\lambda$4686 was absent on optical
spectra taken in 1995 and 1996, and the fluxes in practically all of the
emission lines were significantly reduced. The only lines that
remained relatively strong were the Balmer H\,{\sc i} lines (H$\alpha$
brightened by a factor of $\sim 4$ since 1991; Fig. 6), although weak
low-excitation lines from Fe\,{\sc ii} and forbidden transitions were also
visible. The Balmer decrement was unusually steep, H$\alpha$:H$\beta \sim
30$, and both H$\alpha$ and H$\beta$ had deep central absorptions (Fig. 9).

\begin{figure}
\psfig{file=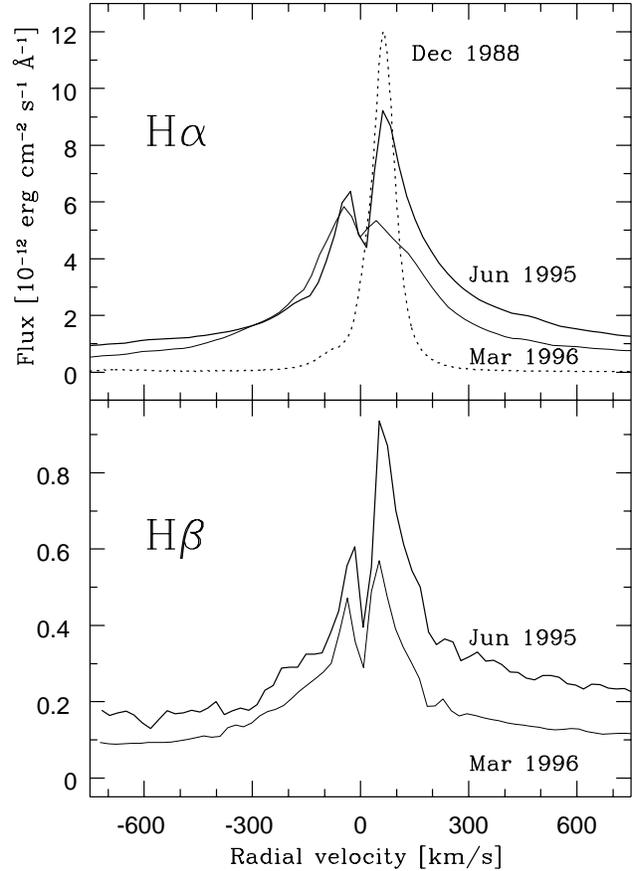,width=0.98\hsize}
\caption{H\,{\sc i} Balmer lines in 1995 and 1996 (thick
and thin line, respectively) exhibited
strong central absorptions which were probably responsible for the
very steep decrement. The H$\alpha$ profile
observed in 1988 (dotted line) by van Winckel et al. (1993) is shown for
comparison. }
\end{figure}

RX~Pup also exhibits interesting variability at radio wavelengths. In the
mid-1970s, when detected for the first time, the radio emission was
characterised by a flat, optically thin spectrum. By 1985 it had changed to
a steeply rising spectrum (Seaquist \& Taylor 1987). In the 1990s, it
flattened at all wavelengths with the previous level of flux density
maintained only at the longest wavelength (IS94). The transition from a
steep to a flat radio spectrum seems to have occurred between February and
December 1991 (IS94). The 3.5 cm flux dropped from $S_{\rm 8.3 GHz} = 69.7
\pm 3.5$ mJy in February 1991 (Seaquist, Krogulec \& Taylor 1993) to $S_{\rm
8.9 GHz} = 11.3 \pm 0.3 $ mJy in July 1992 (ATCA; Lim 1993) and was
apparently related to changes in the UV and optical spectra.

In contrast to the optical brightness, the infrared $JHKL$ light curves are
dominated by periodic variations consistent with those expected for a Mira
variable, and a long term, wavelength dependent, decrease of the mean
brightness (Fig. 4). A Fourier analysis of the data in Table 4 confirms the
pulsation period of 578 day found by Feast et al. (1977). The following
ephemeris gives the phases of minima in the $J$ band:
 \begin{equation}
{\rm Min}(J) = 2\,440\,214 + 578 \times E
\end{equation}
  The long term trends were attributed to dust obscuration possibly related
to the orbital motion (W84; Whitelock 1987). The IR fading is strongest in
the $J$ band, and decreases in amplitude with increasing wavelength. We
estimate the mean fading rates in 1979-1991, $\Delta J \sim 0.2$, $\Delta H
\sim 0.12$, and $\Delta K \sim 0.08\, \rm mag\,yr^{-1}$, respectively. The
fading started in late 1979, following the decline in visual brightness and
appearance of the high-excitation emission line spectrum. The faintest
infrared flux was measured around 1991, after which the $JHKL$
magnitudes recovered to the values observed in 1982-1983.

Several authors claimed that the visibility of molecular bands from the Mira
is limited to the periods of low excitation (IS94, and references therein).
In fact, although RX~Pup was undoubtedly in the low excitation phase in the
1970s, red and blue spectra taken in 1972 show no trace of absorption bands
(Swings \& Klutz 1976). Similarly, Swings \& Klutz failed to detect any TiO
bands at $\lambda \la 8600 \rm \AA$ on their spectra taken in March 1975.
Strong TiO bands in the $\lambda\lambda 7589-8432$ range as well as VO
$\lambda$7900 were detected, for the first time, on November 30, 1980
(Andrillat 1982), after RX~Pup entered a high excitation phase (!). Although
Schulte-Ladbeck (1988) did not find any TiO at $\lambda \la 8000 \rm
\AA$, the TiO as well as VO bands were visible beyond $\lambda \sim 8000 \rm
\AA$ in April 1984 (AW88). Similarly, there were no detectable absorption
bands on our red spectra taken in 1990 and 1991 ($\lambda \la 7200 \rm
\AA$), while the absorption bands beyond $\lambda \sim 7200 \rm \AA$ were
present on spectra taken in 1993 by IS94, and on our spectra taken in 1995 and
1996. In all cases, the absorption bands of the Mira were detected beyond
$\lambda \sim 7000 \rm \AA$. In particular, the TiO bands at $\lambda 6180
\rm \AA$ and $\lambda 7100 \rm \AA$, strong in most symbiotic systems (Kenyon
\& Fernandez-Castro 1987), were never detected in RX~Pup, including our well
exposed spectra of 1995 and 1996. The only positive detections were made in
the long wavelength range, $\lambda \ga 7200 \rm \AA$, when the visual
magnitude of RX~Pup was $V\sim m_{\rm vis} \ga 10.5$ and near maxima of the
Mira's pulsations. The permanent absence of the molecular absorption bands
in the optical range is consistent with the fact that the Mira pulsations
were never visible in that range (Fig. 4). Similar behaviour is shown by
other symbiotic Miras. In particular, Mira pulsations were never observed in
the optical light of HM Sge, even during quiescence (Yudin et al. 1994).

The evolution of the $B-V$ and $U-B$ colour indexes in RX~Pup also suggest
that the contribution of the Mira to the optical light was never
significant. The $B-V$ index had roughly constant value $\sim 0.9$ in
1975-1990, which increased to $\sim 1.7$ in 1993, while the $U-B$ index
increased from $\sim 0.1$ in 1975 to $\sim 0.5$ in 1979, then decreased to
$\sim -0.5$ in 1989-1990, and increased again to $\sim 0.4-0.5$ in 1993
(Feast et al. 1977; Klutz 1979; Munari et al. 1992; IS94). AW88 found the
1984 optical continuum resembled reddened gaseous emission. We estimate
contribution of the emission lines to the observed broad band $B$ and $V$
magnitudes, $\Delta V \sim 1/0.25/0.05$ and $\Delta B \sim 1.8/0.45/0.05$
mag in 1990/91/95, respectively.
During the 1993 observation the Mira component was near maximum of its
pulsation variability ($\phi \sim 0.9$). Given the reddening towards RX~Pup,
E$_{B-V} \ga 0.6$ (Section 3.2), and that near maximum light long period
Miras have $B-V \sim 1.6$ and $U-B \sim 1$ (Smak 1964), the observed values
of $U-B$ and $B-V$ for RX~Pup are similar to those found in other symbiotic
systems, and indicate the hot component dominates at these wavelengths.

Fig. 10 summarises the spectral development of RX~Pup during the last
three decades. The parallel between RX~Pup and the symbiotic novae AG~Peg
(Gallagher et al. 1979; Kenyon et al. 1993), V1329 Cyg (Kenyon 1986, and
references therein), HM Sge (Nussbaumer \& Vogel 1990) as well as the
symbiotic recurrent nova RS~Oph (Kenyon 1986) is apparent, although these
other systems seem to be evolving on very different time scales.

\begin{figure}
\psfig{file=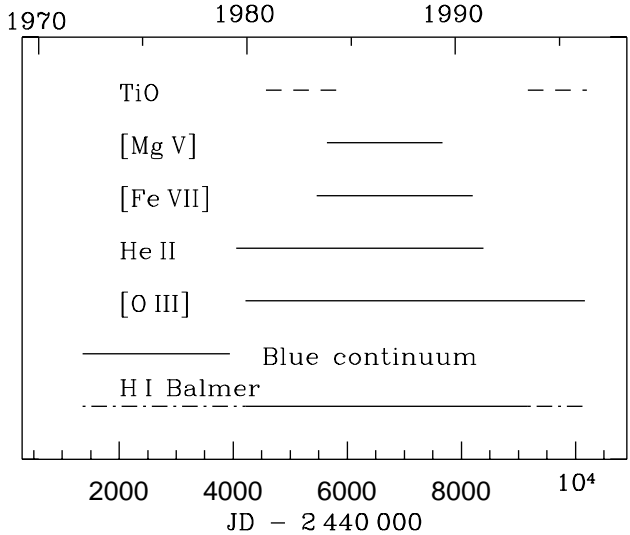,width=0.98\hsize}
\caption{Spectral development of the 1969-1990 outburst of RX~Pup.
Solid, dashed and dot-dashed lines indicate emission and absorption features,
and complex P Cyg profiles, respectively.}
\end{figure}

\subsection{Reddening and distance}

Before deriving the physical characteristics of the RX~Pup binary, we need a
good estimate for the extinction and distance.  We base our reddening
estimates on the UV continuum, UV and optical emission line ratios, IR
photometry, and measurements of the interstellar features. We also assume
the reddening to be represented by a standard interstellar extinction curve
(e.g. Seaton 1979).

\subsubsection{Interstellar absorption features}

RX~Pup lies in field 99(256/--3.0) of Neckel \& Klare (1980), for which they
estimate the interstellar reddening to be E$_{B-V} \sim 0.6$ at $d \sim 1.3
- 2$ kpc.  For such an E$_{B-V}$ we would expect measurable interstellar
features.

In fact, Klutz, Simonetto \& Swings (1978) reported interstellar Na\,{\sc i}
$\rm D_1, D_2$ absorption lines as well as broad absorption bands at
$\lambda \sim 5780$ and 5797 $\rm \AA$. The total equivalent width, $\rm
EW(D_1+D_2) \sim 0.43 {\rm \AA}$, measured by Klutz et al. implies E$_{B-V}
\la 0.1$ (Sembach, Danks \& Savage 1993; Munari \& Zwitter 1997), while
EW$(\lambda\,5780) = 0.45 \rm \AA$ and EW$(\lambda\,5797) = 0.52 \rm \AA$
combined with calibrations of Josafatsson \& Snow (1987) and Herbig (1993)
indicate a much higher E$_{B-V} \sim 0.8 - 1.0$. The latter value is
comparable to our estimates based on emission line ratios (see below), but
higher than the interstellar reddening predicted by Neckel \& Klare (1980).

The absorption lines of Na\,{\sc i} $\rm D_1, D_2$ are also clearly visible
in our optical spectra taken in 1995 and 1996 (Fig. 8). The profiles are,
however, very complex, with faint emission features truncated by very strong
absorption components, $\rm EW(D_1) \sim EW(D_2) \sim 1 {\rm \AA}$, at $\sim
-2$ km/s, and weaker components, $\rm EW(D_1) \sim 0.3-0.4$ and $\rm EW(D_2)
\sim 0.1 {\rm \AA}$, at $\sim -180$ km/s, on both spectra. Traces of both
absorption components are also visible in the emission line profiles of
Na\,{\sc i} on the spectrum taken in April 1991. The ESO spectrum (May 1991)
has insufficient resolution to show them. In addition, the 1995 spectrum
shows strong K\,{\sc i} $\lambda\lambda$\,7665,\,7699 absorption lines [$\rm
EW(7665) \sim 1.3 {\rm \AA}$, $\rm EW(7699) \sim 0.8 {\rm \AA}$] at $\sim 1$
km/s, very close to the radial velocity of the stronger components of the
Na\,{\sc i} D lines. It seems unlikely that these strong
absorption features are of interstellar origin. The radial velocities of the
K\,{\sc i} absorption lines and of the stronger components of the Na\,{\sc
i} lines are $\sim-16$ km/s and $-19$ km/s, respectively, with respect to
the local standard of rest (LSR) and they are blueshifted with respect to
the velocity predicted by galactic rotation law ($v \sim 5-10$ km/s for a
reasonable range $d \sim 1-2$ kpc). The K\,{\sc i} line is somewhat
asymmetric, while the Na\,{\sc i} absorption lines seems to consist of 2
components. The strong absorption components of the Na\,{\sc i} doublet as
well as the K\,{\sc i} absorption lines are probably superpositions of the
interstellar and circumstellar/stellar components.

Adopting the calibration of Bromage \& Nandy (1973), we estimate from the
combined equivalent widths of both Na\,{\sc i} D lines ($\rm
EW(D_1+D_2) that \sim 2 \AA$) E$_{B-V} \ga 0.6$. Other calibrations give similar
results, for example Munari \& Zwitter (1997) find $\rm EW(D_1) \sim 1 \AA$
for E$_{B-V} \ga 0.5$ and stars with multi-component interstellar lines.
However, the K\,{\sc i} absorption lines are too strong for any plausible
E$_{B-V}$ (e.g. Chaffee \& White 1982; Munari \& Zwitter 1997).

The spectra taken in 1995 and 1996 also show the diffuse interstellar bands
at $\lambda \sim$ 5780, 5797 and 6284 $\rm \AA$. Their equivalent widths,
$\rm EW(\lambda 5780) \approx 0.21 \AA$, $\rm EW(\lambda 5797) \approx 0.12
\AA$, and $\rm EW(\lambda 6284) \approx 0.22 \AA$, are consistent with
E$_{B-V} \sim 0.5$ (Herbig 1993; Josafatsson \& Snow 1987; Snow, York \&
Welty 1977).

\subsubsection{UV continuum}

An independent reddening estimate towards the hot component of RX~Pup can be
derived from the intensity of the $\lambda2200$ band. Unfortunately,
practically all {\it IUE} spectra have very poorly exposed continua. The only
spectra that can be used are SWP~2684 and LWR~2395, taken near the visual
maximum in 1978. These show characteristic features of the iron curtain
absorption and a strong absorption band around $\lambda 2200 \rm \AA$. We
compared the combined SWP + LWR spectrum with the {\it IUE} spectra of
CH Cyg taken during its 1981--1984 visual maximum (Miko{\l}ajewska et al.
1988), and estimated E$_{B-V} \la 0.6$.

The very strong absorption feature near $\lambda 2200 \rm \AA$ is also
visible on the HST spectrum taken in May 1991. Correcting this spectrum for
E$_{B-V} \sim 0.8 \pm 0.1$ entirely removes the feature, and the dereddened
spectrum resembles that of CH Cyg observed during the 1985 minimum
(Miko{\l}ajewska et al. 1988). Given the uncertainty in the shape of the hot
component UV continuum, these estimates generally agree with those derived
from the emission lines.

\subsubsection{Emission lines}
 Both the [Ne\,{\sc v}] $\lambda\lambda$\,1575,\,2975 lines, originate from
the same upper level, as do the [Fe\,{\sc vii}]
$\lambda\lambda$\,5721,\,6086 lines. Thus their intensity ratio should
depend only on the transition probabilities, [Ne\,{\sc
v}]\,$I(\lambda1575)/I(\lambda2975)= 1.5$, and [Fe\,{\sc
vii}]\,$I(\lambda5721)/I(\lambda6086) = 0.65$, respectively. The average
[Ne\,{\sc v}] ratio, $\sim 0.6$ measured on the {\it IUE} spectra taken in
1983--1988, implies E$_{B-V} = 0.4\pm 0.4$. The high uncertainty being due
to the faintness of the [Ne\,{\sc v}] lines. The [Fe\,{\sc vii}] line ratio
observed in 1983 by W84 indicates E$_{B-V} = 1.4\pm 0.5$, and data from
FPC92 give E$_{B-V} = 1.9\pm 0.5$ in 1989.
These values should be however considered as upper limits for E$_{B-V}$
because the [Fe\,{\sc vii}] $\lambda 6086$ line can be strongly blended
with the [Ca\,{\sc v}] $\lambda 6086$ line.

A better reddening estimate can be made from the [Ne\,{\sc iv}]
$\lambda$1602/$\lambda$4725 ratio (e.g. Schmid \& Schild 1990a). The
intensity of the ultraviolet [Ne\,{\sc iv}] $\lambda$1602 emission line was
practically constant in 1980-89. We compared its average flux with that in
the optical [Ne\,{\sc iv}] $\lambda$4725 line (W84), and estimate $E_{\rm
B-V} = 0.9 \pm 0.1$.

The [O\,{\sc ii}] $I(\lambda7325)/I(\lambda2471)$ ratio has a practically
constant value of 1.27 for electron temperatures, $T_{\rm e}$, between 5\,000
and 20\,000 K and densities, $n_{\rm e}$, from 10 to $10^6\, \rm cm^{-3}$
(Cant{\'o} et al. 1980). The $\lambda\, 2471$ line is absent on the {\it IUE}
spectra taken before 1983, while a very weak emission appears on later
spectra. The optical [O\,{\sc ii}] $\lambda$7325 doublet was measured on our
1995 and 1996 spectra, and it seems to be also present on the 1993 spectrum
(IS94). The $\lambda\, 7325$ emission was also present on the 1984 spectrum
published by AW88, and on the 1989 spectrum measured by FPC92. Our estimate
of $I(\lambda7325)/I(\lambda2471) \ga 40$ in 1984 and 1989, is consistent
with E$_{B-V} \ga 0.7$. Similarly, $I(\lambda7325)/I(\lambda2471) \ga 13$ in
1993-1996, suggests E$_{B-V} \ga 0.5$.

Extinction parameters can also be estimated from optical H\,{\sc i},
He\,{\sc i} and He\,{\sc ii} recombination lines. The ratios of He\,{\sc ii}
$\lambda$1640, $\lambda$2733 and $\lambda$3203 lines observed in 1982--1988
are consistent with case B recombination and $E_{\rm B-V} = 0.7 \pm 0.1$.
The He\,{\sc ii} $\lambda$1640 emission line flux was practically constant
in 1982--1984. Combining the average flux in the UV line with that in 
He\,{\sc ii} $\lambda$4686 measured in 1983 by W84, we estimate
$I(\lambda1640)/I(\lambda4686) \approx 0.3$ on JD 2\,445\,459, which
indicates $E_{\rm B-V} \approx 0.8$. Finally, the published He\,{\sc ii}
$\lambda$4686/$\lambda$5411 ratios indicate E$_{B-V} \sim 1.4$ (W84), and
$\sim 1.2$ (FPC92), respectively. These values should be, however, considered
as upper limits because the $\lambda5411$ line is blended with Fe\,{\sc ii}
$\lambda5413$, and [Fe\,{\sc iv}] $\lambda5426$.

The H\,{\sc i} Balmer lines are reliable reddening indicators only for
negligible self-absorption in the lower series members (case B
recombination, e.g. Cox \& Mathews 1969; Netzer 1975). For most D-type
symbiotic systems reddening values estimated assuming case B recombination
agree with those derived using other methods (Miko{\l}ajewska, Acker \&
Stenholm 1997). In the case of practically all S-type systems, and some
D-types, there are significant departures from case B, and the
reddening-free ${\rm H\alpha}/{\rm H\beta}$ ratio is $\sim 5-10$ (e.g.
Miko{\l}ajewska \& Kenyon 1992a; Miko{\l}ajewska et al. 1995; Proga et al.
1996; Miko{\l}ajewska et al. 1997) due to self-absorption. In particular,
H$\alpha$/H$\beta \sim 5 -10$ is found at the high electron densities,
$n_{\rm e} \sim 10^9 - 10^{10} {\rm cm^{-3}}$ (Drake \& Ulrich 1980) which
characterise the Balmer line formation region (see Section 3.4).

The values of H$\alpha$/H$\beta$, H$\gamma$/H$\beta$ and H$\delta$/H$\beta$
from W84 and FPC92 are inconsistent with case B recombination for any
reddening. If the ${\rm H\alpha}$ optical depth is $\tau_{\rm H\alpha} \sim
5 - 10$, the lines ratios imply E$_{B-V} \sim 1$. The H\,{\sc i} ratios
measured in 1991 (JD\,2\,448\,385; Table 3) are consistent with case B
recombination, and E$_{B-V} = 1.0 \pm 0.1$. Finally, our observations in
1995 and 1996 show a very steep Balmer decrement, H$\alpha$/H$\beta = 24$
and 31, respectively. Similarly, IS94 measured H$\alpha$/H$\beta = 31$ in
June 1993. IS94 also derived E$_{B-V} \sim 0.4$, comparing the flux in
H$\beta$ with the radio flux at 5 GHz observed at the same epoch. However,
their estimate assumes case B recombination, and is entirely inconsistent
with the E$_{B-V} \sim 2.3$ suggested by the H$\alpha$/H$\beta$ ratio. The
H\,{\sc i} emission line profiles observed in 1995 and 1996 are complex with
central reversals (see also Fig. 9). The $\rm H\alpha$, $\rm H\beta$ and
$\rm H\gamma$ lines are apparently affected by self-absorption. The higher
members are generally less affected. In particular, the H$\gamma$/H$\beta$
ratio differs from the case B value by less than 10 per cent as long as
$\tau_{\rm H\alpha} \la 100$ (Netzer 1975; Proga et al. 1996). Assuming that
the bulk of H\,{\sc i} Balmer emission originates from the same region as
the optical He\,{\sc i} and UV intercombination lines, we estimate
$\tau_{\rm H\alpha} \la 20$ in the 1980s and 1995-96, and $\la 0.5$ in 1991.
The observed H$\gamma$/H$\beta$ ratio was practically constant in 1983-1996
within the observation accuracy of 5-10 per cent, and its mean value $0.32
\pm 0.01$ is consistent with E$_{B-V} = 0.8 \pm 0.3$.

The calculations of Proga, Miko{\l}ajewska \& Kenyon (1994) demonstrated
that the He\,{\sc i} $I(\lambda6678)/I(\lambda5876)$ emission line ratio
distinguishes between S- and D-type systems in that it is $\approx 0.25$ for
D-types and $ \ga 0.5$ for S-types. The average He\,{\sc i}
$I(\lambda6678)/I(\lambda5876) = 0.38 \pm 0.03$ for RX~Pup indicates
E$_{B-V} = 1.0 \pm 0.3$.

The emission line ratios thus generally indicate the reddening
of E$_{B-V} \sim 0.8$ to 1.0 during the period 1980-1991.

\subsubsection{Mira component}

It is clear (see Section 3.1) that the Mira component made a negligible
contribution to the optical light, at least during the last three decades.
Therefore, estimates of the reddening toward the Mira must be based on an
analysis of the infrared. Since the IR light curves (Fig. 4) shows large
variations in the average light and colour of the Mira, we made independent
estimates for 4 different epochs corresponding to different stages of the
hot component (see also Section 3.3):

\begin{description}
\item[{epoch I:}]
JD~2\,442\,767-4\,000: the optical maximum and low excitation phase.
\item[{epoch II:}]
JD~2\,445\,000-6\,600: constant hot component luminosity and high
excitation phase.
\item[{epoch III:}]
JD~2\,447\,500-8\,500: optical and IR minimum.
\item[{epoch IV:}]
JD $\ga$ 2\,448\,500: quiescence and low ionisation.
\end{description}

The colours of Miras have been discussed by Feast et al. (1989) for the LMC,
Whitelock et al. (1994) for the South Galactic Cap (SGC) and Glass et al.
(1995) for the Bulge\footnote{Note that these papers present data on two
different photometric systems, thus transformations must be made before
stellar colours can be compared}. These results suggest that the colours of
the Miras depend to some extent on the environment in which they occur,
probably as a result of the different metallicities of the various
environments.  However, $(J-K)$ appears to be less sensitive to metallicity
variations than are other colours. It also provides a longer baseline for
reddening estimates than $(J-H)$ or $(H-K)$ and avoids any contribution from
dust emission which may effect the $L$ observations.  We therefore use
$(J-K)$ in this analysis in preference to other IR colours.

The reddening towards the Mira was estimated using the $(J-K)$ period-colour
relation derived by Whitelock et al. (1998 in preparation) for Miras in the
solar neighbourhood: \begin{equation} 
(J-K)_0= -0.78+ 0.87 \log {\rm P}
\end{equation}
 Thus for RX~Pup, $(J-K)_0=1.63$. The values derived are listed in table 5.

W83 suggested that there may have been a hot source contributing to the IR
during what we are now calling phase I. We also note that the Mira pulsation
amplitude was low ($\Delta J \sim 1$, $\Delta K \sim 0.8$ in epoch I, and
$\Delta J \sim 1.4$, $\Delta K \sim 1$ during remaining epochs) at this
stage - consistent with the additional contribution from a warm
non-pulsating source. We therefore deduce that during phase I, the hot
component contributed significantly to the $J$ flux.  The value of E$_{B-V}$
tabulated for this phase is therefore likely to be an underestimate. A
similar effect may have been observed in HM Sge, where the amplitude of Mira
pulsations decreased from $\Delta\,J \sim 1.2$ at the visual maximum during
the hot component outburst to $\Delta\,J \sim 2$ mag during the late
outburst decline.

Unfortunately, almost all well-studied long period Miras ($P \ga 400^{\rm
d}$) experience circumstellar reddening, although some have much more than
others (e.g. Whitelock et al. 1994; Glass et al. 1995). Three Miras in the
SGC have periods similar to RX~Pup, they are AW Psc, AFGL\,5592 and
NSV\,14540 (Whitelock et al. 1994). The colours of these stars ($(J-K)$ =
1.9 to 2.4) may be presumed to be unaffected by interstellar reddening.
Thus, if we assume interstellar (or interstellar plus circumbinary)
reddening produces E$_{J-K}\sim 0.4$ (corresponding to E$_{B-V}=0.8$), it
appears that the colours of RX~Pup were quite normal during phases II and
IV.  During phase III, however, the Mira was considerably redder.

\begin{table}
\caption{Reddening towards the Mira component of RX~Pup}
\label{}
\begin{tabular}{lccl}
\hline

\multicolumn{1}{l}{Epoch} & $\overline{K}$ & $\overline{J-K}$ & 
\multicolumn{1}{c}{E$_{B-V}$} \cr
& \multicolumn{3}{c}{(mag)}\\
\hline
I   & 2.37 & 2.13 & $>1.0$ \cr
II  & 2.90 & 2.72 & 2.3 \cr
III & 3.68 & 3.24 & 3.3 \cr
IV  & 3.12 & 2.58 & 2.0 \cr
\hline
\end{tabular}
\end{table}

We also find that the IR continuum energy distribution of RX~Pup observed by
{\it IRAS} (during epoch II) is practically identical to the IR continua
of the long period SGC Miras (AW Psc, AFGL 5592 and NSV 14540) mentioned
above. This supports the interpretation that, during phases II and IV, the
Mira in RX~Pup was a normal high-mass-loss AGB star. Whitelock et al. (1994)
also show correlations between mass-loss rates and $K-[12]$ colour,
pulsation period, and the amplitude of the $K$-light curve for their SGC
sample. The observed $K-[12] \approx 4.6$ (the $K$ magnitude has been
corrected for E$_{B-V} \sim 1$) for RX~Pup implies $\dot{M}_{\rm c} \sim 4
\times 10^{-6}\, \rm M_{\sun}\,yr^{-1}$, and is similar to the values
suggested by the pulsation period and the $K$ amplitude.

The very high reddening value obtained for phase III supports our suggestion
(Section 3.1 and 4.3) that the decline in the IR flux was due to obscuration of
the Mira by dust. A much shorter phase of increased reddening also occured
in 1983/84. The possible causes of these events will be discussed later,
however, the fact that the emission line ratios and the UV continuum do not
indicate any increase in the reddening, implies that the site of the
obscuration must be in the vicinity of the Mira leaving the hot component
and the ionised region outside.

After $\sim$ JD~2\,445\,000 the $K$ magnitude of RX~Pup does not seem to be
much affected by the hot component radiation, and can be used to estimate a
the distance via the period luminosity relation (Feast et al. 1989; van
Leeuwen et al. 1997):
 \begin{equation}
M_{\rm K} = 0.88 - 3.47  \log P.
\end{equation}
 Thus $M_{\rm K} = -8.7 $, although we note that this relation is not well
calibrated for long period Miras. The average $K$ magnitude corrected for
the respective values of E$_{B-V}$ did not change in epoch II-IV; $<K_{\rm
0}> = 2.64$, and implies a distance of $d = 1.8 \pm 0.5 {\rm kpc}$.

In summary, the above analysis is consistent with constant reddening,
E$_{B-V} \sim 0.8\pm0.2$ mag, towards the hot component and the ionised
nebular region, with an additional variable reddening of E$_{B-V} \sim 0$ to
2.5 mag towards the Mira. In the following analyses of the hot component and
the ionised nebula values of E$_{B-V} = 0.8$ and $d = 1.8$ kpc are adopted.

\subsection{Temperature and luminosity of the hot component}

Three methods are used to determine the temperature, $T_{\rm h}$, of the hot
component. Wherever possible, $T_{\rm h}$ is estimated from the observed
equivalent width of He\,{\sc ii} $\lambda$1640 emission line from $IUE$
spectra, and by the He\,{\sc ii}\,$\lambda$4686/H$\beta$ or He\,{\sc
ii}\,$\lambda$4686/He\,{\sc i}\,$\lambda$5876 ratios from optical spectra.
Since H$\beta$ may be weakened by optical depth effects, we assume
He\,{\sc ii}\,$\lambda$4686/He\,{\sc i}\,$\lambda$5876 is a more reliable
indicator of the temperature than He\,{\sc ii}\,$\lambda$4686/H$\beta$ for
the period 1983-1990, when the optical depth effects seem to be important.
The third method applies the relation between $T_{\rm h}$ and the highest
observed ionisation potential, $\chi_{\rm max}$,
 \begin{equation}
T_{\rm h}~[1000\,{\rm K}] = \chi_{\rm max}~[\rm eV],
\end{equation}
 proposed by M{\"u}rset \& Nussbaumer (1994). The accuracy of this method is
about 10 per cent provided that the highest ionisation stage is indeed
observable. The highest observed ionisation stages as well as the derived
temperatures at various epochs are presented in Table 6. The highest
ionisation stage observed in $IUE$ spectra of RX~Pup is that of Mg$^{+4}$,
and in optical spectra are those of Fe$^{+6}$ and O$^{+5}$; they suggest
temperatures in good agreement with estimates based on other methods.

To derive the luminosity of hot component(s), $L_{\rm h}$, the following
methods are used: \begin{description}
 \item[{(a)}] The optical and IR photometry obtained during visual maximum
($\sim $ JD~2\,442\,500 - 4\,000) suggests negligible contribution of the
cool giant to the $UBV$ magnitudes. While the $JHKL$ magnitudes contain
contributions from both the cool and the hot components. We assume that the
IR magnitudes observed in epoch IV (see Section 3.2.4) represent those of
the Mira, and that the contribution of the hot component to the IR continuum
was negligible at that time. To estimate the $JHKL$ magnitudes of the hot
component in remaining epochs, we simply subtracted the contribution of the
Mira from the observed values. The values for $L_{\rm h}$ in the last column
of Table 6, correspond to the integrated (UV)-optical-IR luminosity. We
combined the $UBV$ magnitudes for JD~2\,442\,524 (Feast et al. 1977) with
the $JHKL$ data for JD~2\,442\,520, and the $UBV$ magnitudes for
JD~2\,443\,627 (Klutz 1979), with the $IUE$ spectra taken on JD~2\,443\,770
and the average IR data from epoch I.
 \item[{(b)}]
Our second method is based on the flux
of the He\,{\sc ii}\,$\lambda$1640 emission line measured on $IUE$ spectra,
or the He\,{\sc ii}\,$\lambda$4686, H$\beta$ and H$\alpha$ fluxes derived from
optical spectra.
We assumed a blackbody spectrum for the hot component, and that
H\,{\sc i} and He\,{\sc ii} emission lines are produced
by photoionisation followed by recombination (case B).
The case B assumption is apparently not satisfied for
the H\,{\sc i} emission lines in 1983-1990,
and thus the corresponding values for $L_{\rm h}$ in Table 6
are those derived
from He\,{\sc ii}. Or, for JD\,2\,447\,510, from H$\alpha$, whose flux
should not differ from its case B value by more than 20 per cent
(Netzer 1975).
The He\,{\sc ii}\,$\lambda$1640 fluxes measured on low resolution
{\it IUE} spectra before 1983 have been corrected
for the contribution of the O\,{\sc i}]\,$\lambda$1641 line estimated
for the high resolution spectra. This contribution was
$\sim 40$ per cent of the total (He\,{\sc ii}\,$\lambda$1640
+ O\,{\sc i}]\,$\lambda$1641)
flux in 1980 September, $\sim 30/25$ per cent in 1981 June/1982 March,
and it becomes negligible after 1983 October.
 \item[{(c)}] Miko{\l}ajewska et al. (1997) found a relation between the hot
component luminosity in symbiotic binary systems and the flux emitted in the
Raman scattered O\,{\sc vi}\,$\lambda$6825 emission line. Schmid \& Schild
(1990b) report a weak emission at $\lambda \sim 6830 \rm \AA$ on 1 January
1990, with an equivalent width of $\sim 11.7\, \rm \AA$ which they identify
as O\,{\sc vi} $\lambda$6825. Similarly, Harries \&
Howarth (1996) estimate EW($\lambda$6825) $\sim 0.15 \pm .05\, \rm \AA$ in
May 1994. We converted these equivalent widths to fluxes estimating the
continuum level from published $R_{\rm c}$ photometry (Munari et al. 1992;
IS94). O\,{\sc vi} $\lambda$6825 was below our detection
limit, $\sim 10^{-13}\, \rm erg\,s^{-1}\,cm^{-2}$, on all our optical
spectra, including those taken in November 1990.
 \item[{(d)}] In a few cases $L_{\rm h}$ was derived from $UBV$
magnitudes observed during the nebular phase. The method was proposed by
M{\"u}rset \& Nussbaumer (1994), who also gave the bolometric corrections to
the $UBV$ magnitudes for a wide range of $T_{\rm h}$. For JD~2\,447\,978 and
JD~2\,448\,215/23 published $UBV$ magnitudes (Munari et al. 1992; IS94) were
used, while for JD~2\,448\,328, the $UBV$ magnitudes were calculated from
our HST spectrum.
 \item[{(e)}] The HST spectrum was also used to calculate the integrated
($\lambda\lambda 1770-5450 \rm \AA$) flux. The flux emitted shortwards of
1200 $\rm \AA$, $L_{\rm EUV,h}$, can be estimated from the optical H$\beta$,
He\,{\sc i}\,$\lambda$5876, and He\,{\sc ii}\,$\lambda$4686 emission lines
(Kenyon et al. 1991; Eq.(8)). The sum of the integrated HST flux and $L_{\rm
EUV,h}$ derived from the data in Table 3 for JD~2\,448\,352 and
JD~2\,448\,385 agrees very well with the values indicated by the method (b),
and is by a factor of $\sim 2$ lower than the value derived from $UBV$
magnitudes (method d). \end{description}
   We estimate the luminosities in Table 6 have accuracies of a factor of 2,
as indicated by the values obtained at the same epoch, but using different
methods. In addition, there is a systematic error due to uncertainties in 
reddening and distance estimates; this, however, should not affect any
conclusions about the evolution of the hot component luminosity with time.

The data in Table 6 demonstrate that the hot component had roughly constant
luminosity in 1975-1989, although its temperature changes by a factor of 15.
Between 1989 May and 1991 April, $L_{\rm h}$ dropped by a factor of $\sim
20$, while $T_h$ decreased by $\ga 40$ per cent. The evolution of RX~Pup in
1975-1991 thus parallels the evolution of a typical symbiotic nova (compare
Miko{\l}ajewska \& Kenyon 1992b; Kenyon et al. 1993; M{\"u}rset \&
Nussbaumer 1994).

In 1995-96, the optical continuum is much stronger than in 1991, while the
nebular emission is significantly reduced. There is no evidence for high
ionisation lines. In particular, we do not see He\,{\sc ii}\,$\lambda$4686
in a spectrum taken in 1996, and the upper limit for the equivalent width of
the Raman scattered O\,{\sc vi} $\lambda$6825 emission line,
EW($\lambda$6825) $\la 0.15 \rm \AA$ is comparable to the value measured in
1994 by Harries \& Howarth (1996), and requires $L_{\rm h}$ of order of 100
$\rm L_{\sun}$. On the other hand, RX~Pup was detected by ROSAT in
April/June 1993 (M{\"u}rset, Wolff \& Jordan 1997), although the {\it IUE}
spectrum from May 1993 does not show any evidence for high ionisation
features, nor does the optical spectrum from June 1993 (IS94) which is very
similar to our 1995/96 spectra. The lack of measurable He\,{\sc ii} emission
(for $T_{\rm h} \ga 10^5$ K and $L_{\rm h} \sim 100\,\rm L_{\sun}$ the
observed flux in He\,{\sc ii}\,4686 should be of $\sim 10^{-13}\, \rm
erg\,s^{-1}\,cm^{-2}$) may then be due to obscuration of the He\,{\sc ii}
formation region, which must be located very close to the hot component.

The H\,{\sc i} Balmer emission lines are relatively strong, and their
equivalent widths, EW($\rm H\alpha) \approx 200-250 \AA$ and EW($\rm H\beta)
\approx 16-24 \AA$, measured on our 1995-96 spectra, respectively, are
comparable to those observed in quiescent recurrent novae T CrB and RS~Oph
(Dobrzycka et al. 1996a; Anupama \& Miko{\l}ajewska 1999). 
The He\,{\sc i} emission lines are, however, very
faint or absent. The colour temperature of the optical $\lambda\lambda
4500-5500 \rm \AA$ continuum is low, $T \sim 7000-6000$ K for E$_{B-V} =
0.8$. This component has a luminosity of $L \sim 600 (d/1.8\,{\rm kpc})^2
\rm L_{\sun}$. On the other hand, the optical emission lines indicate a
higher temperature, $T \sim 2.5-4 \times 10^4$ K, with a roughly comparable
luminosity, $L \sim 500 - 1000 (d/1.8\,{\rm kpc})^2 \rm L_{\sun}$. We find
here a striking similarity between RX~Pup and the quiescent symbiotic
recurrent novae RS~Oph and T CrB, as well as the hot component of CH Cyg
during bright phases. In all these systems, the optical data indicate the
presence of relatively cool F/A- (CH Cyg) or A/B-type source (RS~Oph and T
CrB), while the optical emission lines indicate a higher temperature source,
with a roughly comparable luminosity (Miko{\l}ajewska et al. 1988; Dobrzycka
et al. 1996a; Anupama \& Miko{\l}ajewska 1999). 
We will return to this problem in Section 4.

\begin{table}
\caption{Hot component parameters for RX~Pup}
\label{}
\footnotesize
\begin{tabular}{l l c c c }
\hline
JD~24... & Ion & $T_{\rm h}$\,[K]
& $L_{\rm h}$\,[$\rm L_{\sun}$] & Method \cr
\hline
42524 &  & ~~8\,000 & 10700 & a \cr
43627-770 & & ~~9\,000 & 18000 & a \cr
44055 & $\rm He^{+2}$ & ~80\,000 & ~8800 & b \cr
44072 & $\rm He^{+2}$ & ~80\,000 & 13800 & b \cr
44345 & $\rm Ne^{+4}$ & 100\,000 & 15700 & b \cr
44503 & $\rm Ne^{+4}$ & 100\,000 & 17600 & b \cr
44767 & $\rm Ne^{+4}$ & 110\,000 & 17300 & b \cr
45051 & $\rm Ne^{+4}$ & 110\,000 & 18300 & b \cr
45459 & $\rm Fe^{+6}$ & 120\,000 & 17600 & b \cr
45638 & $\rm Mg^{+4}$ & 120\,000 & 15400 & b \cr
45771 & $\rm Mg^{+4}$ & 120\,000 & 15000 & b \cr
46559 & $\rm Mg^{+4}$ & 120\,000 & 13100 & b \cr
47459 & $\rm Mg^{+4}$ & 120\,000 & 10500 & b \cr
47510 &               & 120\,000 & ~7500 & b \cr
47667 & $\rm Mg^{+4}$ & 120\,000 & ~9100 & b \cr
47893 & $\rm O^{+5}$ & 115\,000 & ~3600 & c \cr
47978 &     & 115\,000 & ~3000 & d \cr
48200 & $\rm O^{+5}$ & 115\,000 & ~3600 & b \cr
48215-23 &           & 110\,000 & ~2700 & d \cr
48328  &        & 80\,000 & ~1000 & d \cr
48328-85  &        &         & ~~550 & e \cr
48352 & $\rm He^{+2}$ & ~80\,000 & ~~430 & b \cr
48385 & $\rm He^{+2}$ & ~80\,000 & ~~430 & b \cr
49461 & $\rm O^{+5}$ & 115\,000 & $\la 360$ & c \cr
\hline
\end{tabular}
\end{table}

\subsection{Nebular emission}

The rich emission-line spectrum that has been developing since late 1979,
offers various possibilities for the determination of the nebular parameters
$T_{\rm e}$ and $n_{\rm e}$. Published estimates for the electron density
indicate systematic differences between the values derived from the UV
intercombination lines and those derived from the forbidden optical emission
lines. In particular, K82 estimate $n_{\rm e} \sim 10^{10}\, \rm cm^{-3}$
from {\it IUE} observations in 1981 June, while W84 derive $n_{\rm e} \ga
10^5$ and $\sim 10^7\, \rm cm^{-3}$, from the [S\,{\sc ii}]
$\lambda$6717/$\lambda$6731 and [O\,{\sc iii}]
$\lambda$4363/($\lambda$4959+$\lambda$5007) line intensity ratios,
respectively, observed in 1983. Based on the optical spectrum taken in
December 1989, FPC92 simulate the density gradient inside the ionised region by
adopting a two-layer model, in which different ionisation stages are
present, and derive $n_{\rm e} \sim 4.5 \times 10^6\, \rm cm^{-3}$ from high
excitation [O\,{\sc iii}] and [Fe\,{\sc vi}] emission line ratios, and $n_e
\sim 2.6 \times 10^5\, \rm cm^{-3}$ from low excitation [N\,{\sc ii}] and
[S\,{\sc ii}] line ratios, respectively. All these estimates assume a
photoionised region with $T_e \sim 10^4\, \rm K$.

The significance of the density gradient in the ionised region is also
indicated by the O\,{\sc iii}]/[O\,{\sc iii}] line ratio. Nussbaumer \&
Storey (1979) demonstrated that simultaneous observations of the UV
intercombination O\,{\sc iii}] $\lambda\lambda$1661, 1667 and the optical
forbidden [O\,{\sc iii}] $\lambda\lambda$4363, 4959, 5007 emission lines
enable both $T_e$ and $n_e$ to be determined. Although the $IUE$ and optical
observations presented here are not simultaneous, we can combine the O\,{\sc
iii}] $\lambda$1664 fluxes from Table 2 with published measurements for the
optical [O\,{\sc iii}] lines. In particular, the 1983 [O\,{\sc iii}] fluxes
from W84 combined with the mean of O\,{\sc iii}] $\lambda$1664 flux measured
on the {\it IUE} spectra taken in 1982 and 1983, require a region with $n_e
\sim 10^{9} - 10^{10}\, \rm cm^{-3}$ and $T_e \sim 7500\, \rm K$ for
E$_{B-V} \sim 0.8 - 1$. The O\,{\sc iii}]/[O\,{\sc iii}] intensity observed
in the late 1980s (Table 2; van Winckel et al. 1993; FPC92) also require
high densities and very low temperatures $T_e \la 7000$ K. Such a low 
temperature is, however, in conflict with the absence of strong lines
arising from dielectronic recombination. A similar effect is seen in other
symbiotic systems, including symbiotic novae (e.g. Schmid \& Schild 1990a).
To solve this problem, Schmid \& Schild inferred that the forbidden lines
come from a much more extended region (by a factor of $10^4$ or so) than the
O\,{\sc iii}] emission, both having similar $T_e \ga 10^4$ K.

There is no simple method of determining $T_e$ in symbiotic systems.
Nussbaumer \& Storey (1984) have proposed the use of dielectronic
recombination lines. The idea rests on a comparison of a collisionally
excited line with one produced by dielectronic recombination. Nussbaumer \&
Storey also list promising recombination lines longwards of $912\, \rm \AA$.
The best candidates among these are N\,{\sc iv} $\lambda$1719 and C\,{\sc
ii} $\lambda$1335. The observed flux ratio N\,{\sc v}\,$\lambda$1240/N\,{\sc
iv}\,$\lambda$1719 grew from $\approx 1.5$ in 1981 to $\approx 3$ in
1984, and $\approx 8$ in 1986. The corresponding temperatures are $T_e
\approx 12\,000\, \rm K$ in 1981, 13\,000 K in 1984, and 15\,000 K in 1986,
for E$_{B-V} \sim 0.8$. A very weak C\,{\sc ii}\,$\lambda$1335 line is
present on spectra taken in 1979 and 1980. The observed flux ratios C\,{\sc
iii}]\,$\lambda$1909/C\,{\sc ii}\,$\lambda$1335 $\sim 60$ in 1979, and $\ga
100$ in 1980 indicate $T_e \sim 12\,000 - 13\,000$ K. These temperatures
argue very strongly in favour of a radiatively ionised UV emission line
region.

The line ratios within the UV O\,{\sc iv}]\,$\lambda$1401 and N\,{\sc
iii}]\,$\lambda$1750 multiplets are very sensitive to $n_e$ (Nussbaumer \&
Storey 1979, 1982). The N\,{\sc iii}] $\lambda$1754/$\lambda$1749.7 and
N\,{\sc iii}] $\lambda$1752.2/$\lambda$1749.7 ratios measured on {\it IUE}
high resolution spectra indicate the electron density was decreasing from
$n_e \ga 10^{10} - 10^{11}$ in 1980 to $\sim 5 \times 10^9$ in 1982-83 and
$\sim 10^9\, \rm cm^{-3}$ in 1987. Similarly, we estimate $n_e \sim 10^{10}$
in 1981 and $\sim 10^{9}\, \rm cm^{-3}$ in 1987 from the O\,{\sc iv}]
$\lambda$1407.4/$\lambda$1404.8 ratio. These results are practically
independent of $T_e$. We also observe a decreasing trend in the Si\,{\sc
iii}]\,$\lambda$1892/C\,{\sc iii}]\,$\lambda$1908 and an increasing trend in
the N\,{\sc iii}]\,$\lambda$1750/C\,{\sc iii}]\,$\lambda$1908 line ratios
measured on the low resolution {\it IUE} spectra. These trends can be
interpreted as a continuously decreasing $n_e$. Finally, the increase in the
C\,{\sc iv} doublet ratio in 1980-1988 (see Section 3.1.2) is consistent
with decreasing optical depth (which is related to density) in the
high-excitation permitted line region.

We can now determine whether or not the parameters derived from the
intercombination lines are also representative of the H\,{\sc i}/ He\,{\sc
i} emission region. The ionisation potentials of $\rm O^{+2}$ (54.9 eV) and
$\rm He^+$ (54.4 eV) are very similar, so He\,{\sc i} and O\,{\sc iii}]
should be emitted in the same physical region. The fluxes in He\,{\sc i}
$\lambda$5876 and O\,{\sc iii}] $\lambda$1664 then require $n_e \sim 10^{10}
\, \rm cm^{-3}$ in 1983-84 and $n_e \sim 10^9 \, \rm cm^{-3}$ in 1988/90,
and $T_e \sim 15\,000$ K in 1984-90. The dependence of emissivity in
He\,{\sc i} $\lambda$5876 on $T_e$ and $n_e$ was taken from Proga et al.
(1994) and Proga (1994), while for O\,{\sc iii}] $\lambda$1664 data from
Nussbaumer \& Storey (1981) were used. The volumes needed to account for the
O\,{\sc iii}] $\lambda$1664 and He\,{\sc i} 5876 emission fluxes are $V \sim
8 \times 10^{40}$ and $\sim 2 \times 10^{42}\, (d/1.8\, \rm kpc)^2\, \rm cm^3$ 
in 1983-84, and 1988-90, respectively, so the corresponding radii are $R \sim
2$ and $\sim 5\, \rm a.u.$ The resulting emission measures are a factor of
$\sim 2$ higher than corresponding values derived from the H$\beta$, and
roughly equal those derived from the H$\alpha$ emission fluxes. However,
these estimates assume that the H\,{\sc i} lines follow case B. For a
constant density gas, the optical depth in H$\alpha$ is $\tau_{\rm H\alpha}
\approx 10 (R/1\,{\rm a.u.})/(n_e/10^{10}\,\rm cm^{-3})$ (Cox \& Matthews
1969). We estimate $\tau_{\rm H\alpha} \sim 20$ and $\sim 5$ in 1983-84 and
1988-90, respectively. In such conditions the H$\beta$ flux could be reduced
by $\sim 50-70$ per cent with respect to the case B value, while H$\alpha$
should be somewhat (by $\la$ 20 per cent) enhanced. These results indicate 
that the UV intercombination lines form in the same gas as do the H\,{\sc i}
and He\,{\sc i} lines.

Our spectra taken in 1991 do not show any measurable UV emission lines, while
the optical H\,{\sc i}, He\,{\sc i} emission lines remain strong. The
He\,{\sc i} line ratios are consistent with formation in a region
with $n_e \sim 10^{6} - 10^{8}\, \rm cm^{-3}$ (Proga et al. 1994), while the
H\,{\sc i} line ratios are consistent with case B for E$_{B-V} \sim 0.8 -
1$. We estimate the emission measure $n_e^2V \sim 1.4 \times 10^{59}$ and
$\sim 1.2 \times 10^{59}\,(d/1.8\, \rm kpc)^2\, \rm cm^{-5}$, for the He\,{\sc i}/H\,{\sc
i} line emitting region ($T_{\rm e} \sim 15\,000$ K), and the expected
O\,{\sc iii}] $\lambda$1664 flux of $\la 7 \times 10^{-15}\, \rm
erg\,s^{-1}\,cm^{-2}$ which is below the detection limit on our {\it IUE}
spectrum.

We now examine the low density region in which the forbidden lines are
produced. Based on an analysis of the forbidden lines in three
symbiotic novae Schmid \& Schild (1990a) found that physical conditions vary
strongly throughout the nebulosities. In particular, they found a steep
electron density gradient from the lowest to the highest observed ionisation
stages. Following their suggestion, we have grouped the diagnostic line
ratios used according to the ionisational potential of the ions involved:
 \begin{description}
\item[{(a)}]
the transition region between the ionised and neutral gas,
characterised by [S\,{\sc ii}] and [O\,{\sc i}];
\item[{(b)}]
the $\rm He^+$ zone and the outer edge of the $\rm He^{+2}$
region, described by [N\,{\sc ii}], [Fe\,{\sc iii}], [Ar\,{\sc iii}],
[O\,{\sc iii}] and [Ne\,{\sc iv}]; and
\item[{(c)}]
the inner $\rm He^{+2}$ region, characterised by [Ne\,{\sc v}],
[Mg\,{\sc v}], [Fe\,{\sc vi}] and [Fe\,{\sc vii}].
\end{description}

The [S\,{\sc ii}] $\lambda$6716/$\lambda$6731 ratio is consistent with $n_e
\sim {\rm few} \times 10^4\, \rm cm^{-3}$ and $T_e \sim 10^4$ K (Cant{\'o} et
al. 1980) for the entire observing period. The electron temperature can be
in principle constrained by [O\,{\sc i}]
($\lambda$6300+$\lambda$6364)/$\lambda$5579 (Keenan et al. 1995), and
the data in Table 3 are consistent with $n_e \sim 2\, \times 10^{4}\, \rm
cm^{-3}$ and $T_e \sim 10\,000$ K in 1991, and $n_e \sim 5\, \times 10^{4}\,
\rm cm^{-3}$ and a rather high $T_e \la 18\,000$ K in 1995. However,
[O\,{\sc i}] $\lambda$5579 can be severely affected by the atmospheric
airglow (Keenan et al. 1995) so the [O\,{\sc i}] ratio only constrains an
upper limit for $T_e$.

The strong [Ar\,{\sc iii}] $\lambda$7135 and very weak or absent [Ar\,{\sc
iii}] $\lambda$5192 lines (Table 3; FPC92) also indicate low $T_e$. Our
estimate of $I(\lambda7135)/I(\lambda5192) \ga 100$ from the 1991 spectrum
implies $T_e \la 15\,000$ K for E$_{B-V} \sim 0.8$ (Keenan, Johnson \&
Kingston 1988). Unfortunately, our 1990 spectrum does not cover the $\sim
\lambda\,5200\, \rm \AA$ region. The absence of [Ar\,{\sc iii}]
$\lambda$5192 in Table 2 of FPC92 suggests $T_e \la 20\,000$ K in 1989. The
[N\,{\sc ii}] $\lambda$6584/$\lambda$5755 ratio observed in 1989-91
indicates $n_e \ga 1.5 \times 10^5\, \rm cm^{-3}$ for $T_e \la 15\,000$ K
(Kafatos \& Lynch 1980). Similarly, we estimate $n_e \ga 8 \times 10^6$,
$\ga 4 \times 10^6$, and $\ga 2 \times 10^6\, \rm cm^{-3}$ from [O\,{\sc
iii}] ($\lambda$5007+$\lambda$4959)/$\lambda$4363 observed in 1983 (W84),
1989 (FPC92) and 1991 (Table 3), respectively (Nussbaumer \& Storey 1981),
and $n_e \sim 5 - 9 \times 10^6\, \rm cm^{-3}$ from the [Ne\,{\sc iv}]
$\lambda$1602/$\lambda$2423 ratio measured in 1986-87 (Nussbaumer 1982). The
[Fe\,{\sc iii}] intensities are consistent with $n_e \sim 10^5 - 10^6$ in
1990, and $n_e \sim 10^5\, \rm cm^{-3}$ in 1991, respectively (Keenan et al.
1993). The volume needed to account for $L(\rm [O\,{\sc iii}]\,\lambda5007)$
is $V \sim 1.7 \times 10^{46}$, $\sim 1.6 \times 10^{46}$, and $\sim 6
\times 10^{45}\,(d/\rm 1.8\,kpc)^2\, \rm cm^3$ in 1983, 1988-90 and 1991,
respectively, (Nussbaumer \& Storey 1981), and the corresponding radii are
$R \sim 107/103/75\, (d/\rm 1.8\,kpc)^{2/3}$ a.u. for a spherically
symmetric nebula.

There is also no evidence for high $T_e$ in the region where the emission
lines with the highest ionisation potential originate. [Mg\,{\sc v}]
$\lambda\lambda$2784,\,2930 are very strong on several {\it IUE} spectra
taken in 1986-89; $\lambda\,2417$ is, however, very weak or absent. Our
estimate of $I(\lambda2784+\lambda2930)/I(\lambda2417) \la 30$ is consistent
with $T_e \la 20\,000$ K for $n_e \la 10^8\, \rm cm^{-3}$, and E$_{B-V} \sim
0.8$ (Kafatos \& Lynch 1980). Similarly, the absence of [Ar\,{\sc v}]
$\lambda$4626 on our optical spectrum taken in 1990, implies the observed
ratio [Ar\,{\sc v}] $I(\lambda 6435 + \lambda 7005)/I(\lambda 4626) \la 6$,
and $T_e \la 20\,000$ K for $n_e \la 10^8\, \rm cm^{-3}$ and E$_{B-V} \sim
0.8$. The [Fe\,{\sc vi}] $\lambda$4967/$\lambda$4973 and [Fe\,{\sc vii}]
$\lambda$4967/$\lambda$4989 ratios measured on our 1990 spectrum are then
consistent with $n_e \sim 10^6 - 10^7\, \rm cm^{-3}$ (Nussbaumer \& Storey
1978; 1982). The volume needed to account for the luminosity in the
[Fe\,{\sc vii}] $\lambda\lambda$6086,\,5721 lines is then comparable to that
required for [O\,{\sc iii}] lines if Fe/H is similar to that found in other
symbiotic novae and M giants (e.g. Schmid \& Schild 1990a).

The physical conditions in the nebula are thus consistent with those
expected from a photoionised gas. The bulk of permitted and intercombination
lines are formed in a dense region of $\sim$ a few a.u. and $n_e \sim 10^9 -
10^{10}\, \rm cm^{-3}$, with an average $T_e \sim 15\,000$ K. The forbidden
lines come from more extended region of $\sim 100$ a.u. ([O\,{\sc iii}],
[Fe\,{\sc vii}]) or larger. We also find a steep density gradient covering
more than 2 orders of magnitudes from the highest ($\rm Fe^{+6}$) to the
lowest ($\rm S^+$) observed ionisation stage in the forbidden line emitting
region. $T_e$ also decreases with decreasing ionisation. The conditions in
the nebula were evolving during the period under discussion. The average
$T_e$ was increasing with increasing ionisation level till c. 1989, then in
1991 it decreased in step with the ionisation degree. The opposite trend is
observed in $n_e$; it was slowly decreasing till c. 1989. In both the
permitted/intercombination line emitting region and in the forbidden line
region the average $n_e$ decreased by an order of magnitude. This trend is
also manifested by a general increase in the forbidden emission line fluxes
with respect to the permitted lines with similar ionisation potentials.

A comparison of the conditions derived from the H\,{\sc i}, He\,{\sc i} and
O\,{\sc iii}] lines with those derived from the forbidden lines, [O\,{\sc
iii}] and [Fe\,{\sc vii}], suggests the density gradient observed in RX~Pup
is roughly of a form $n_e \sim r^{-2}$, which is consistent with the
emission from an ionised wind emanating from the system. A similar
conclusion is reached from an analysis of the radio observations (Hollis et
al. 1986; Seaquist \& Taylor 1987, 1992). This shows the spectrum and
angular structure expected from partially optically thick thermal
bremsstrahlung originating in a spherically symmetric steady stellar wind.
The cm-mm spectrum observed in 1985-1988 is essentially linear from 1.49 to
394 GHz, and is well represented by a spectrum of the form $S_{\rm \nu} =
9.9\, \nu^{0.80}\, \rm mJy$ (Seaquist \& Taylor 1992). The photospheric
radius at 394 GHz is given by $r \sim 29\, (d/1.8 \rm kpc)$ a.u., and the
turnover frequency $\nu_{\rm t} \ga 394$ GHz implies $n_e \ga 8 \times
10^7\, \rm cm^{-3}$ inside that radius.

For a spherically symmetric steady wind emanating predominantly either from
the cool giant or the hot companion, we have:
 \begin{equation}
n_e = \frac{\dot{M}}{4 \pi \mu m_{\rm H} v r^2},
\end{equation}
 where $r$ is the radial distance from the centre of the star, $\mu$ the
mean atomic weight of the gas and $m_{\rm H}$ the mass of a hydrogen atom.
Using the characteristic radii and densities derived above from the emission
lines and radio mm data for 1985-88, we estimate $\dot{M}/v \sim 6 \times
10^{-6}\, [\rm M_{\odot}\,yr^{-1}/100\,km\,s^{-1}]$. FW's analysis of
emission line profiles indicate $v \sim 140\, \rm km\,s^{-1}$, and $\dot{M}
\sim 10^{-5}\,\rm M_{\odot}\,yr^{-1}$. It is unlikely that the cool giant
wind has a velocity in excess of 100 $\rm km\,s^{-1}$. We thus assume the
wind originates from the hot star. The evolution in the nebular conditions
can then reflect the changes in that wind. Note that our estimate for
$\dot{M}$, although acceptable for a symbiotic nova (e.g. Kato 1997), is
very approximate. In particular, there is evidence for elongated structure
in the radio and in the optical emission lines (Hollis et al. 1986; Paresce
1990), which indicates the wind may be bipolar.  The observed radio spectral
index of $\sim$ +0.8 also departs from the canonical value of +0.6 expected
for a spherically symmetric wind.

The change in the emission spectrum in 1991 can be understood in terms of
decreased $\dot{M}$ and gradual recombination in the ionised region
following the decline in the hot component luminosity, which is no longer
able to maintain the ionisation of the wind. As the recombination time
scale, $\tau_{\rm rec}$, is inversely proportional to $n_e$ -- $\tau_{\rm
rec} \sim 1\,\rm hr$/1\,month/15\,yr for $n_e \sim 10^9/10^6/10^4\, \rm
cm^{-3}$, respectively -- the inner portions of the winds are first to
recombine. In fact, spectra taken in 1995 and 1996 show low ionisation
emission lines of [O\,{\sc i}], [S\,{\sc ii}], [N\,{\sc ii}] with almost the
same (within a factor of $\sim 2$) intensity as observed in 1990 and 1991,
and very faint [O\,{\sc iii}] lines; all these lines have intensities
roughly consistent with their originating in a region with $n_e \sim 10^4\,
\rm cm^{-3}$ and $T_e \la 10\,000$ K. Such a scenario is also consistent
with the observed changes in radio emission. This suggested an optically
thick wind emanating from the system in the 1980s; while since c. 1991 it is
better understood as optically thin emission (IS94, and references therein)
arising from the low density remnant of the 1980-90 high-ionisation phase.

Only the Balmer H\,{\sc i} and Fe\,{\sc ii} emission lines show increased
intensity on our 1995/96 spectra. This increase coincides with the
appearance of an A/F-type continuum source. The very steep Balmer decrement
is consistent with its origin in a relatively dense region, $n_e \ga 10^{10}\,
\rm cm^{-3}$ (Drake \& Ulrich 1980). Assuming $T_e \la 10^4$ K, we estimate
the optically thin emission measure $n_{\rm e}^2V \sim 4 \times 10^{59}
(d/\rm 1.8\, kpc)^2\, \rm cm^{-3}$ from the H$\beta$ and H$\alpha$ emission
line fluxes. Again, we emphasise here the similarity to the spectroscopic
appearance of CH~Cyg during its active phases. In particular, the luminosity
of the A/F-type continuum, as well as the conditions (density, emission
measure) in the H\,{\sc ii} Balmer emission region at maximum of CH Cyg's
activity (Miko{\l}ajewska et al. 1987, 1988) are comparable to those derived
from the 1995/96 spectra of RX~Pup.

\section{Discussion}

\subsection{Geometrical configuration of RX~Pup}

The large difference between the reddening towards the Mira and the
reddening towards to the hot component and ionised region indicates the hot
component and most of the ionised material must lie along a line of sight
that does not include the Mira variable and its associated dust cocoon.

Kenyon, Fernandez-Castro \& Stencel (1988) deduced a characteristic dust
radius of $\sim 200$ a.u. from {\it IRAS} colours, while Anandarao, Taylor \&
Pottasch (1988) derived a dust shell radius of $\sim 77$ a.u. (re-scaled for
our adopted distance of 1.8 kpc) from analysis of the {\it IRAS} photometry
and low resolution spectroscopy. Both values are an order of magnitude
larger than the binary separation, $a \sim 17$ a.u., suggested by the IS94
model.

Schmid \& Schild (1990b) report weak emission at $\lambda\,6825\, \rm \AA$ in
1990 January due to Raman scattering of the O\,{\sc vi} $\lambda\,1032$
resonance line. They also noted that the lack of detectable polarization in
this line indicates the line of sight coincides roughly with the direction
defined by the $\rm O^{+5}$ zone and the Mira variable. RX~Pup shows at the
same epoch very strong [Fe\,{\sc vii}] emission lines, and the lack of
strongly scattered O\,{\sc vi} $\lambda\,6825$ emission is unusual.
Practically {\it all} symbiotic systems with strong [Fe\,{\sc vii}] lines
have also strong O\,{\sc vi} $\lambda\lambda\,6825,\,7082$ lines (Kenyon
1986; Miko{\l}ajewska et al. 1997). The apparent weakness of the Raman
scattered O\,{\sc vi} line suggests that the scattering process in the
envelope of RX~Pup must be very inefficient, probably due to insufficient
numbers of the scatterers -- neutral hydrogen atoms. Theoretical simulations
of the Raman scattering in symbiotic systems demonstrated that the process
has the highest efficiency in systems having ionisation geometry with an
{\it X}-parameter (as defined by Seaquist, Taylor \& Button 1984, hereafter
STB84), $X \sim 1$, while in systems with $X \ga 5$, the efficiency is
much lower (Schmid 1996). The weak O\,{\sc vi} $\lambda\,6825$ would thus
suggest that in RX~Pup only a cone-shaped region shielded by dense material
surrounding the Mira remains neutral, while the lack of measurable
polarization indicates that in 1990 the $\rm O^{+5}$ zone (and the hot
component) was in front of this neutral region. Moreover, the lack of
changes in the reddening towards the hot component and the nebular region
(Section 3.2) shows that at least during the last 18 years both the hot
component and the emission line region have remained outside the Mira's dust
shell. In particular, this rules out the possibility that the changes in the
extinction towards the Mira component are orbitally related (we return to
this problem in Section 4.3).

The conical shape of the interface between neutral and ionised regions in RX
Pup can be produced either by illumination of the Mira's wind by the hot
companion (STB84) or by interacting winds (Girard \& Willson 1987). We have
already demonstrated that the nebular emission in RX~Pup originates from an
ionised wind emanating from the system, and that the hot component is the
most likely source of this wind. One can expect that this wind will interact
with the Mira's wind. Assuming that the shock front coincides with the
interface between the ionised and neutral regions, then $X \ga 5$ (see
above) corresponds to a momentum ratio, $mv = \dot{M}_{\rm h}v_{\rm
h}/\dot{M}_{\rm c}v_{\rm c} \ga 10$. Adopting $\dot{M}_{\rm h} \sim
10^{-5}\, \rm M_{\sun}\,yr^{-1}$, $v_{\rm h} \sim 140\, \rm km\,s^{-1}$
(Section 3.4), $\dot{M}_{\rm c} \sim 4 \times 10^{-6}\, \rm M_{\sun}\,yr^{-1}$
(Section 3.2.4), and assuming reasonable $v_{\rm c} \sim 20\, \rm km\,s^{-1}$
(Whitelock et al. 1994), we estimate $mv \sim 18$, in agreement with the
suggested geometry. The distance of the shock from the Mira normalised to
the binary separation will then be $\sim 0.2$. The latter result combined
with the fact that the Mira component is never stripped of its dust envelope
(Table 5) implies a binary separation that is a factor of $\sim 5$ or so
larger than the dust formation radius. Assuming a typical dust formation
radius of $\ga 5 \times R_{\rm c}$, and the Mira radius, $R_{\rm c} \sim 2 -
3$ a.u. (e.g. Haniff, Scholz \& Tuthill 1995), the required binary
separation is $a \ga 50$ a.u.

With so large a binary separation, the Mira in RX~Pup should have SiO masers,
whereas the search for such emission in 1987 (43 GHz; Allen et al. 1989),
1989 and 1994 (86 GHz; Schwarz et al. 1995) resulted in an upper limit which
was a factor of $\sim 100$ below fluxes in single Mira variables with
similar pulsation periods. The popular explanations for the apparent absence
of SiO masers (as well as OH and H$_2$O) in most symbiotic systems involve
radiation or tidal effects related to the hot compact companion (Schwarz et
al. 1995, and references therein). These are unlikely to apply to RX~Pup,
mostly because of the large binary separation, but also because a
significant amount of the circumstellar dust is always present outside the
SiO formation region (if the SiO molecules are destroyed by hot radiation
the dust should be destroyed too). Moreover, at least the 1994 observation
was made after the luminosity and temperature of the companion significantly
decreased.

Allen et al. (1989) noted that if the cool giant were always viewed through
the ionised nebula, then the masers would be obscured at frequencies for
which the nebula is optically thick. Moreover, Nyman \& Olofsson (1986)
found a correlation of the 87 GHz flux with the Mira pulsation phase. In
fact, these two effects can explain the failure to detect SiO masers in RX
Pup. First, our analysis suggests the Mira was always viewed through the
ionised nebula. Thus in 1987, when the optically thick radio emission up to
$\ga 300$ GHz was observed (IS94) the 43 GHz maser would indeed be obscured.
It seems that the radio spectrum remained optically thick until c. 1991
(IS94), which indicates that in 1989 the 87 GHz maser could be also have
been obscured. Although the nebula was optically thin at 87 GHz at least in
1994, both the 1989 and 1994 observations were made near the Mira minimum
($\phi \approx 0.42$ and $\phi \approx 0.48$, respectively). The lack of
detectable SiO masers in RX~Pup is therefore consistent with the suggested
geometrical configuration of this system. It also seems reasonable to repeat
the search for such emission near to the Mira maximum, and when the system
is in a low excitation stage with a flat radio spectrum.

\subsection{Evolution of the hot component}

Fig. 11 shows the evolution of the hot component in the
Hertzsprung-Russell diagram -- for the luminosities and temperatures derived
in Section 3.3 -- along with the cooling curves for 0.64 $\rm M_{\sun}$ and
0.55 $\rm M_{\sun}$ white dwarfs, respectively. The hot component appears to
maintain a constant bolometric luminosity, $<L_{\rm h}> = 15\,000 \pm 1\,000
\rm L_{\sun}$, from 1975 to 1986, though its temperature varies by a factor
$\ga 15$.  In 1988/89 it turned over in the HR diagram, since then our
estimates suggest a decline in luminosity initially at roughly constant
temperature. Between May 1989 and April 1991, $L_{\rm h}$ dropped by a
factor of $\sim 20$, while $T_{\rm h}$ decreased by $\ga 40$ per cent. The
evolution of RX Pup during its recent active phase thus parallels the
evolution of a typical symbiotic nova pointing to a thermonuclear
nova eruption (compare Miko{\l}ajewska \& Kenyon
1992b; Kenyon et al. 1993; M{\"u}rset \& Nussbaumer 1994), and in particular
it rules out the accretion model proposed by K85, with its recent
developments by IS94. Neither can the accretion model account for the
emission line profiles or the evolution of the radio spectrum (see Section 3.4
for details). The evolution of $L_h$ is also inconsistent with the
thermonuclear-runaway model proposed by AW88, which assumes a very slow
hydrogen-shell flash with a plateau phase lasting decades. 
In particular,
the AW88 model does not predict any significant changes in $L_h$, as all
photometric and spectral changes have been attributed to changes in the mass
loss from the white dwarf.

The estimated maximum bolometric luminosity, $L_{\rm h} \approx 15\,000 \rm
L_{\sun}$ falls within the range of plateau luminosities for other symbiotic
novae (Miko{\l}ajewska \& Kenyon 1992b; M{\"u}rset \& Nussbaumer 1994).
Applying the relationship between plateau luminosity and white dwarf mass
for accreting cold white dwarfs from Iben \& Tutukov (1996, their equation
8), we estimate the mass of the white dwarf in RX~Pup, $m_{\rm WD} \sim 0.59
\rm M_{\sun}$, while the Paczy{\'n}ski-Uus relation (Paczy{\'n}ski 1970; Uus
1970) gives $m_{\rm WD} \sim 0.77 \rm M_{\sun}$. The luminosity-mass
relation during the plateau phase, however, depends significantly on the
thermal history of the white dwarf: generally hot white dwarfs must have
larger masses than the cool ones to reach the same luminosity during a
hydrogen-burning phase (Iben \& Tutukov 1996). Our estimates for the white
dwarf mass of RX~Pup should thus be considered as lower limits.

\begin{figure}
\psfig{file=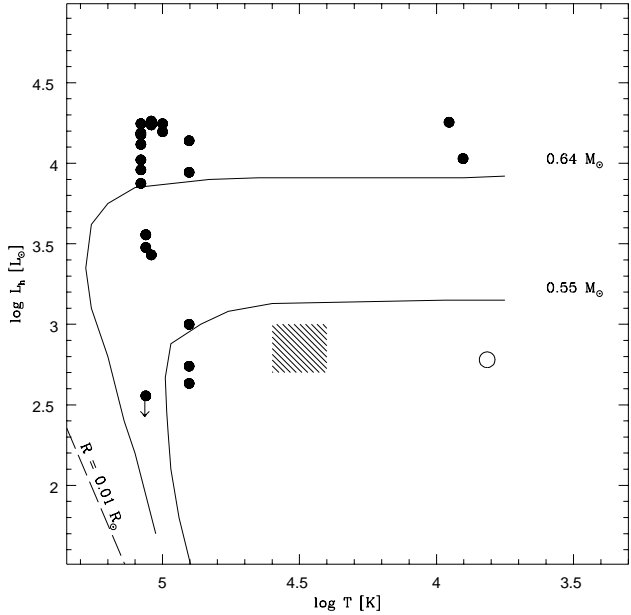,width=0.98\hsize}
\caption{Evolution of the hot component in the HR diagram
in 1975-1996. Dots indicate the 1975-1994 outburst cycle.
The shaded square represents the results derived from H\,{\sc i}
emission lines, while the open circle locates
the A/F-type component observed in 1995-96.
The solid curves are the evolutionary tracks from Sch{\"o}nberner (1989),
the dashed curve is a line of constant radius.}
\end{figure}

RX~Pup may be the first recurrent nova detected in a D-type symbiotic
system. Published observations suggest RX~Pup was visually bright in 1894,
and its spectrum resembled that of $\eta$ Car, with strong H\,{\sc i} Balmer
emission lines (Pickering 1897). Three years later a high excitation nebular
spectrum emerged, with strong emission lines of He\,{\sc ii}, [O\,{\sc
iii}], [Ne\,{\sc iii}], in addition to the H\,{\sc i} Balmer series (Cannon
1916). After 1900, RX~Pup entered a deep optical minimum ($m_{\rm pg} =
14.1$ mag in 1904-5; Pickering 1914) resembling that in 1991. The star
recovered from the minimum in 1909, and from then up until 1924 it showed
only small fluctuations around $m_{\rm pg} \sim 12$ mag (Yamamoto 1924). The
high excitation spectrum was again observed in 1941 (Swings \& Struve 1941),
while in 1949-51 the excitation level was moderate and the optical continuum
was very weak (Henize 1976). In the late 1960s RX~Pup remained very faint
optically ($m_{\rm vis} = 12.5, m_{\rm pg} = 13.5$) and the high excitation
features were absent (Sanduleak \& Stephenson 1973). In 1969/70 the $V$
magnitude brightened to 9.4 mag (Eggen 1973), and the recent outburst began.
The historical records of RX~Pup thus suggest that around 1894 it underwent
another nova-like eruption.  Unfortunately, the lack of observations in the
1930s makes it impossible for us to ascertain whether the presence of a
high-excitation spectrum in 1941 was also due to the eruption.

The recurrence time for RX~Pup, $\tau_{\rm rec} \sim 80 (40)$ yr, is
comparable to that of two symbiotic recurrent novae, T CrB (80
yr) and RS~Oph (22 yr), and it requires an accretion rate of the order of
$\sim 10^{-7} {\rm M_{\sun}\,yr^{-1}}$ (Prialnik \& Kovetz 1995). According
to Bondi-Hoyle theory even in a wide binary system a white dwarf can accrete
as much as $\sim 1$ per cent of the giant's stellar wind (Livio 1988, and
references therein), which in the case of RX~Pup is just what is needed. The
observed quiescent luminosities of the compact components of Mira AB ($\sim
2\, \rm L_{\sun}$; Karovska et al. 1997)\footnote{Their original values are
scaled to the photometric parallax, 118 pc (e.g. Van Leeuwen et al. 1997),
which is in good agreement with the Hipparcos trigonometric parallax.}
and R Aqr (a few $\rm L_{\sun}$;
Michalitsianos, Kafatos \& Hobbs 1980) spatially resolved wind-accreting
binary systems agree with the values predicted by the accretion theory. The
component separation of Mira AB, $ a \sim$ 70 a.u. (Karovska et al. 1997) is
comparable to that expected for RX~Pup, while the quiescent $L_{\rm h}$ in
RX~Pup is a factor of $\sim 100$ higher than that of Mira B in good
agreement with more or less the same order of magnitude difference in the
mass-loss rates of the Mira in RX~Pup ($\dot{M}_{\rm c} \sim 4 \times
10^{-6}\, \rm M_{\sun}\,yr^{-1}$; Section 3.2.4) and that of Mira A
($\dot{M}_{\rm c} \sim 10^{-7}\, \rm M_{\sun}\,yr^{-1}$; Reimers \&
Cassatella 1985).

The observed maximum luminosity as well as the very slow evolution in the
H-R diagram indicate a white dwarf mass significantly less than the $M_{\rm
h} \sim 1.4\, \rm M_{\sun}$ usually expected for recurrent novae
(e.g. Webbink et al. 1987). As the
estimated $L_h$ is based largely on UV emission line fluxes, it will be
underestimated if the nebula is not ionisation bounded. However, any
underestimate is unlikely to be sufficient to increase the white dwarf mass
above $\sim 1\,\rm M_{\sun}$. Moreover, recent model calculations show that
while recurrent novae require high mass-accretion rates, $\dot{M} \ga
10^{-8}\, \rm M_{\sun}\,yr^{-1}$, the constraint on the white dwarf mass is
less severe (Prialnik \& Kovetz 1995). Short recurrence times can be  obtained
for white dwarf masses as low as $\sim 1 \rm M_{\sun}$. Consequently, the
decline times of recurrent nova models are not necessarily short. Given all
the uncertainties involved in the theoretical modelling of symbiotic and
recurrent novae (Sion 1997, and references therein), it seems plausible to
suggest that RX~Pup is a recurrent nova.

The quiescent hot component of RX~Pup shows striking similarities with its
counterparts in RS~Oph and T~CrB. In particular, the ROSAT spectrum of RX~Pup
observed in April-June 1993 (M{\"u}rset et al. 1997) is very similar to the
ROSAT spectrum of RS~Oph (Orio 1993); both show $\beta$ type spectra
(according to the classification of M{\"u}rset et al.) compatible with
emission from an optically thin plasma with a temperature of a few $10^6$~K.
M{\"u}rset et al. suggest that such emission is due to hot shocked gas, and
propose that the shock arises in the region where the strong and fast wind
from the hot component crashes into the cool giant or its wind. Such an
interpretation is, however, entirely incompatible with the observed
properties of both RX~Pup and RS~Oph at the time of the ROSAT observations.
The colliding wind model requires very $L_{\rm h} \sim 10\,000 \rm L_{\sun}$
for RX~Pup, and $\sim 2\,000 \rm L_{\sun}$ for RS~Oph, which is much higher
that the observed $L_{\rm h} \la 1000 \rm L_{\sun}$ in RX~Pup, and $\sim 100
- 600 L_{\sun}$ in RS~Oph (Dobrzycka et al. 1996a). Moreover, the optical
spectrum of RS~Oph at that time was very similar to the quiescent (1993-96)
spectrum of RX~Pup, showing only weak low-ionisation and low-excitation
emission lines (Dobrzycka et al. 1996a).

It seems thus reasonable to assume (as did Orio for RS~Oph) that the X-ray
emission arises in the accretion flow. Assuming that the ROSAT flux
represents a significant fraction of the total flux in the 0.2-4 KeV range,
a mass accretion rate $\dot{M}_{acc} \approx 10^{-9} (M_{\rm WD}/0.7\,{\rm
M_{\sun}})^{-1.8} {\rm M_{\sun}\,yr^{-1}}$ (Patterson \& Raymond 1985) can
be estimated. However, this should be considered a lower limit to
$\dot{M}_{acc}$, because the X-ray luminosity for RX~Pup given by M{\"u}rset
et al. was derived assuming a much lower reddening, E$_{B-V} = 0.55$, than
that derived above. In addition, local material might absorb a substantial
fraction of the X-rays produced close to the white dwarf surface.

The quiescent optical spectrum of RX~Pup, in particular the presence of a
variable relatively cool F-type continuum with strong H\,{\sc i} Balmer
emission lines, also resembles the optical spectra of RS~Oph and T~CrB.
These stars show an additional variable component in the optical-UV and
sometimes also at IR wavelengths. For example, in RS~Oph the IR colours
indicate the presence of an additional warm, $\ga 7000$ K source (Evans et al.
1988), while Dobrzycka et al. (1996a) found an A-B type shell source with $L
\sim 100 - 600 \rm L_{\sun}$, accompanied by strong H\,{\sc i} and moderate
He\,{\sc i} emission lines. Similarly, Selvelli, Cassatella \& Gilmozzi
(1992; also Belczy{\'n}ski \& Miko{\l}ajewska 1998;
Anupama \& Miko{\l}ajewska 1999) reported variable
UV/optical continuum with $L \sim 40-100 \rm L_{\sun}$ in T~CrB. Although
both in the symbiotic recurrent novae, T~CrB and RS~Oph, and in RX~Pup,
the average luminosity of the B/A/F-type shell source is consistent with the
accretion rate, $\dot{M} \ga 10^{-8}\, \rm M_{\sun}\,yr^{-1}$, required by
the theoretical models, the effective temperatures places the hot components
far from the standard massive white dwarf tracks in the HR diagram.
Simultaneously, the X-ray data suggest 1-2 orders of magnitude lower
accretion rates. All three systems show similar quiescent behaviour: their
hot components have highly variable luminosity and occasionally display
blue-shifted absorption features (see also Dobrzycka et al. 1996a;
Anupama \& Miko{\l}ajewska 1999).

It is also interesting that the shell spectrum appears only during the late
decline from the nova outburst. In particular, the optical/visual light
curves from the outbursts of T~CrB, RS~Oph and RX~Pup show more or less
pronounced minima followed by a standstill (or secondary maximum) associated
with the appearance of the shell spectrum. We suggest the minima are due to
a decline in the hot component's luminosity after it passes the turnover in
the HR diagram. The strong hot component wind during the plateau phase
prevents accretion onto the hot component. Following the decline in
luminosity the wind also ceases, and accretion of the material from the cool
giant can be restored. We believe that the shell-type features and variable
``false atmosphere" observed at quiescence together with the H\,{\sc i}
emission line spectrum originate from the accretion flow. The observed
variability could be due to fluctuations in the mass-loss rate from the cool
giant, the shell spectrum becomes stronger as $\dot{M}_{\rm c}$ increases
and as a result $\dot{M}_{\rm acc}$ increases. Finally, we recall the
behaviour of CH~Cyg, a highly variable symbiotic system containing an
accreting white dwarf. The hot component luminosity in this system varies by
a factor of $10^4$ (Miko{\l}ajewska 1994), between $\sim 0.1$ and $\sim
300\, \rm L_{\sun}$. The brightening of the hot component is associated with
the presence of a flickering, which suggests the system is
accretion-powered. The optical/UV spectrum of CH Cyg during bright phases
resembles the quiescent spectra of the symbiotic recurrent novae (see also
Section 3.3), while the detection of hard X-ray emission 
(M{\"u}rset et al. 1997; Ezuka, Ishida \& Makino 1998) indicates 
that the accreting component is a white dwarf.

\subsection{Obscuration of the Mira component}

\begin{figure}
\psfig{file=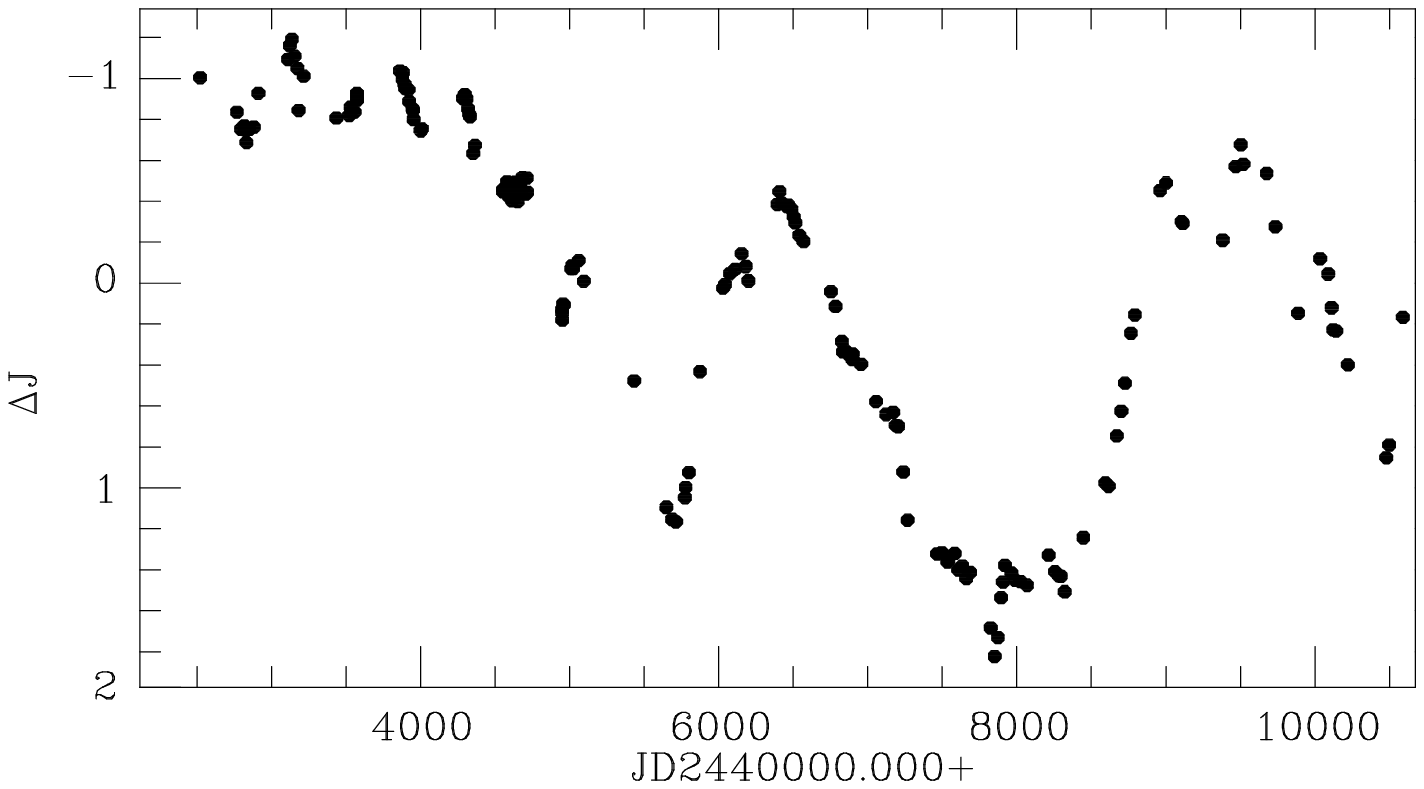,width=0.98\hsize}
\psfig{file=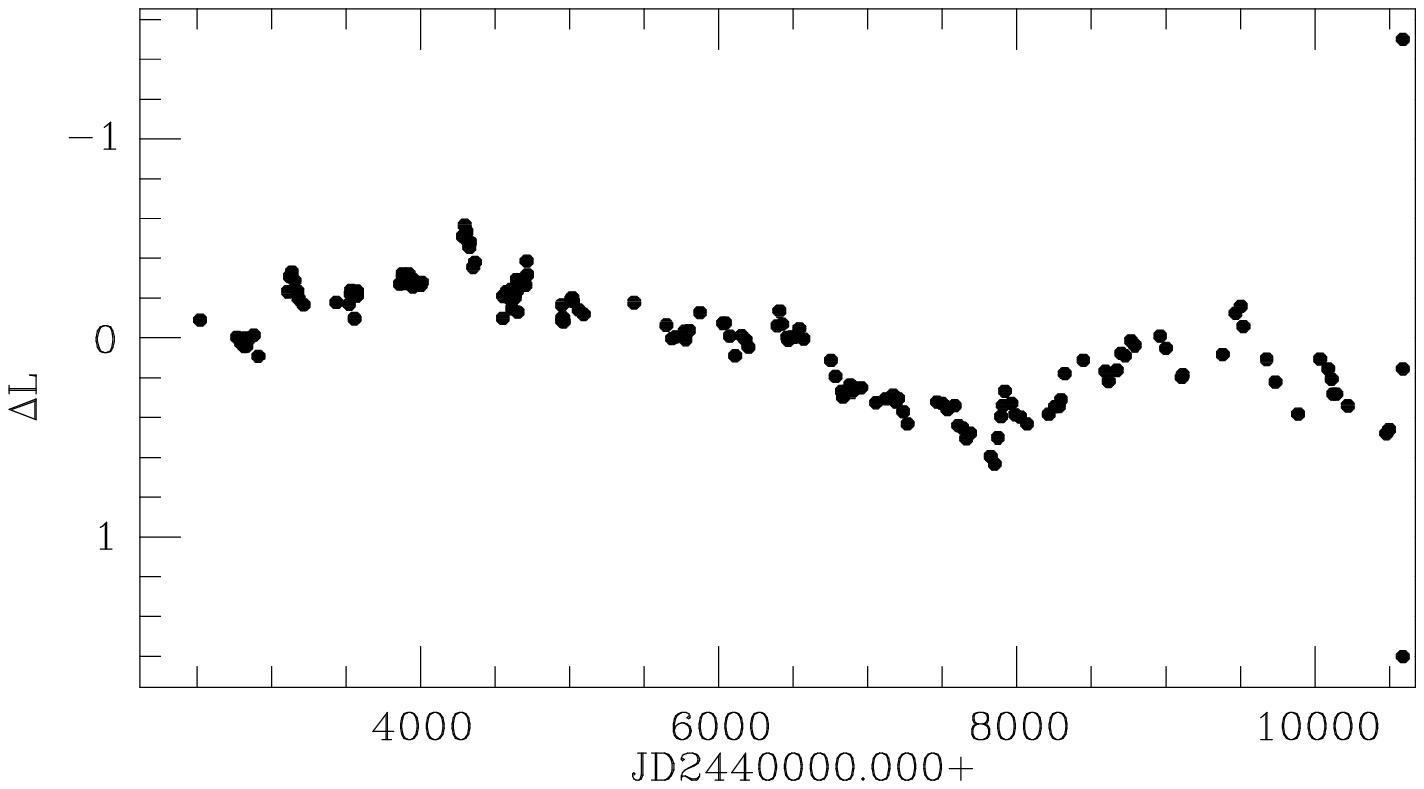,width=0.98\hsize}
\caption{$J$ and $L$ light curves after the removal of the Mira
pulsation.}
\end{figure}

The light curves of symbiotic Miras were discussed by Whitelock (1987,
1988). RX Pup is typical of these in that it shows significant long term
variations (refereed to as ``obscuration events" by Whitelock) in addition
to the Mira pulsation. Fig. 12 shows the $J$ and $L$ light curves after
removal of the Mira pulsations. This is done by subtracting the best fitting
578 day sine-curve from the data as displayed in Table 4 and Fig. 4. Only $J$
and $L$ are illustrated as they represent the extremes of behaviour.
During the first approximately 2000 days low amplitude 578 day variations
can be seen in anti-phase with the pulsation cycle; these are artifacts of
the pulsation removal process introduced by the lower amplitude of the Mira
at that stage (see Section 3.2.4).  The $\Delta J$ curve changes over a full
range of about 2.5 mag. While the $\Delta L$ light curve 
shows a much lower amplitude, $ \sim 0.8$ mag, trend, with
a minimum at the same time as the $\Delta J$ curve.

Although the light curves in Fig. 12 seem qualitatively similar to the
visual light curve (Fig. 4), the optical minimum around JD 2\,448\,300 can be
accounted for by the evolution of the hot component, in particular, the
general decline in the emission line spectrum (the change in emission line
fluxes alone can account for a $\sim 1/1.5$ mag drop in $V/B$ magnitudes,
respectively; Section 3.1) following the decrease in its luminosity. In fact,
the visual magnitudes are generally not correlated with the IR magnitude,
although the hot component may contribute to the IR before $\sim$
JD 2\,444\,000 (epoch I/optical maximum; see Section 3.2.4). In particular, the
visual light curve was practically flat during JD 2\,445\,000-8\,200, while
the IR light curves showed pronounced changes (Fig. 12). Similarly, the
decrease in IR flux after $\sim$~JD\,2\,449\,500 is not accompanied by
any related changes in the visual. This, and the fact that the changes in
the reddening towards the Mira are {\it not} correlated with similar changes
in the reddening towards the hot component and emission line formation
region(s), strongly suggest the obscuration affects only the Mira.

The physical nature of this phenomenon remains a mystery. In the specific
case of the symbiotic star R Aqr the long term trends were explained as the
effect of an orbitally related eclipse of the Mira by pre-existing dust
(Willson et al. 1981; Whitelock et al. 1983b). However, more recently
Whitelock et al. (1997) have explained large amplitude erratic variations in
the light curves of carbon-rich Miras as the consequence of the ejection of
puffs of dust by the star. The wavelength dependence of these two effects:
an eclipse by pre-existing dust and the ejection of new dust, are very
similar and it is difficult to determine which is occurring in RX~Pup. From
Fig. 12 it is clear that any eclipsing dust cloud would have to be very
patchy, much more so than that responsible for the eclipse in R Aqr. 

Fig. 12 suggests a modulation with a time scale of $\sim 3000$ days, while
the reddening towards the hot component was practically constant over at
least 13 years. It is hard to imagine a geometry of the binary system in
which the cool component is eclipsed at least twice while the hot component
is not eclipsed at all. Thus, for RX~Pup, it seem unlikely that the
obscuration is orbitally related. Now that data are available over a longer
time base it seems that the obscuration events occur in many symbiotic Miras
with too great a frequency to be generally associated with binary eclipses
(Whitelock 1998).

Although the most pronounced obscuration event coincides with the hot
component passing the turnover in the HR diagram the obscuration is apparently
not associated with this component. In particular, the dust does not form
in the material ejected from the hot component. The relative intensities of
Si\,{\sc iv} $\lambda\,1394$ + O\,{\sc iv}] $\lambda\,1403$/N\,{\sc iv}
$\lambda\,1487$ and O\,{\sc iii}] $\lambda\,1664$/N\,{\sc iii}
$\lambda\,1750$ should not vary with changes in $T_{\rm h}$, $n_{\rm e}$ and
ionisation. The {\it IUE} data in Table 2 show that in RX~Pup these ratios
are roughly constant until at least JD~2\,447\,667 which means that at that
time O and Si are not being depleted due to condensation into dust grains
(silicates). Similarly, there is not indication for carbon depletion in the
ejecta. Moreover, the absence of strong O\,{\sc i} during the late outburst
phase and the steady decline in He\,{\sc ii} emission suggest the hot
component ejecta lack a neutral zone. Similarly, Schmid \& Schild (1990a) do
not find evidence for significant depletion of Si and Fe in the ionised ejecta
of V1016 Cyg and HM Sge, while Munari \& Whitelock (1989) did not find any
major changes of emission line strengths associated with the dust
obscuration episode in HM Sge.

The site of the dust responsible for the symbiotic Mira obscuration must be
located in its circumstellar envelope. In a considerable number of Mira
variables, the visual light curves exhibit complex variations from cycle to
cycle as well as on long time scales. Although the long term modulations are
more evident in C-rich Miras (e.g. R For, RU Vir), they are also present in
the light curves of longer ($\ga 300^{\rm d}$) period O-rich Miras (S CMi, R
Hya; Mattei 1997). Longer period Miras generally have more dust and it 
seems likely that the long term modulation of their light curves is related
to the behaviour of this dust. Because they have more dust they are visually
fainter and therefore less well studied at short wavelengths. It is
therefore possible that these obscuration events have been easily recognised 
in symbiotic Miras simply because their light curves have been more
extensively studied. Whitelock's (1998) discussion suggests this is not
the case, but more extensive studies are necessary to confirm this.

It is not at all clear how, or if, the obscuration events affect the rest of
the symbiotic system. It is possible that an enhancement in the Mira wind
produces a coordinated increase in the accretion luminosity of the hot
component as suggested by the rise in the optical continuum and H\,{\sc i}
emission lines after $\sim$ JD 2\,450\,000 (Section 3.3). A similar relation
between the obscuration of the Mira (and related enhancement in its wind)
and an increase in the hot component luminosity is probably observed in R
Aqr (Miko{\l}ajewska \& Kenyon 1992b).

\section{Conclusions}

RX~Pup is a long-period interacting binary system consisting of a Mira
variable pulsating with $P \approx 578$ days, and a white dwarf companion.
The binary separation could be as large as $a \ga 50$ a.u. (corresponding to
$P_{\rm orb} \ga 200$ yr) as suggested by the permanent presence of a dust
shell around the Mira component. In particular, the Mira is never stripped
of its dust envelope contrary to the model proposed by IS94, and even during
bright (not obscured) IR phases the star resembles the high-mass loss
galactic Miras with thick dust shells.

The analysis of multifrequency observations shows that most, if not all,
photometric and spectroscopic activity of RX~Pup in the UV, optical and
radio range is due to activity of the hot component, while the Mira variable
and its circumstellar environment is responsible for practically all changes
in the IR range. In particular, the evolution of the hot component in the HR
diagram as well as evolution of the nebular emission in 1970-1993 is
consistent with a symbiotic nova eruption, with the luminosity plateau
reached in 1972/75 and a turnover in 1988/89. The hot component contracted
in radius at roughly constant luminosity from c. 1972 to 1986; during this
phase it was the source of a strong stellar wind and therefore could not
accrete any further material. By 1991 the luminosity of the nova remnant had
decreased to a few per cent of the maximum (plateau) luminosity, and the hot
wind had practically ceased. By 1995 the hot component start to accrete
material from the Mira wind, as indicated by a general increase of the
optical continuum and Balmer H\,{\sc i} emission. The quiescent optical
spectrum of RX~Pup resembles the quiescent spectra of symbiotic recurrent
novae, while the hot component luminosity is consistent with variable
wind-accretion at a high rate, $\dot{M}_{\rm acc} \sim 10^{-7}\, \rm
M_{\sun}\,yr^{-1}$ ($\approx 1$ per cent of $\dot{M}_{\rm c}$). RX~Pup may
be a recurrent nova; there is some evidence that a previous eruption
occurred around 1894. The proposed binary model for RX~Pup could be verified
if flickering of the accretion disk were detected at quiescence. Such
flickering has been observed in the symbiotic recurrent novae RS~Oph and T
CrB during quiescence, and in the accretion-powered symbiotic systems CH Cyg
and MWC 560 (e.g. Dobrzycka, Kenyon \& Milone 1996b).

Large changes were found in the reddening towards the Mira, these were
accompanied by fading of the near IR flux. However, the reddening towards
the hot component and emission line regions remained practically constant
and was generally less than that towards the Mira. These changes do not seem
related to the orbital configuration nor with the hot component activity.
The IR dust obscuration episodes are best explained as intrinsic changes in
the circumstellar environment of Mira possibly due to variable mass-loss
rate.

\section*{Acknowledgements}
We gratefully acknowledge the very helpful comments on this project 
by W. Dziembowski, A. Nota, F. Paresce and R. Szczerba.  
We thank the variable star section of the Royal Astronomical
Society of New Zealand for providing us with their visual
magnitudes. We also thank the
following people for making IR observations: Michael Feast, Brian Carter,
Greg Roberts, Robin Catchpole and Dave Laney.
This research was partly supported by KBN Research Grant No. 2 P03D 021 12.

{}

\bsp

\label{lastpage}


\begin{thebibliography}{}

\bibitem[1988]{aw88}
Allen, D.A., Wright, A.E., 1988, MNRAS, 232, 683 (AW88)

\bibitem[]{}
Allen, D.A., Hall, P.J., Norris, R.P., Troup, E.R., Wark, R.M.,
Wright, A.E., 1989, MNRAS, 236, 363

\bibitem[]{}
Anandarao, B.G., Taylor, A.R., Pottasch, S.R., 1988, A\&A, 203, 361

\bibitem[]{}
Andrillat, Y., 1982, in The Nature of Symbiotic Stars,
eds. M. Friedjung \& R. Viotti, D. Reidel, p.47

\bibitem[]{}
Anupama, G.C., Miko{\l}ajewska, J., 1999, A\&A, in press, astro-ph/9812432

\bibitem[1984]{bs84}
Baldwin, J.A., Stone, R.P.S., 1984, MNRAS, 206, 241

\bibitem[]{}
Barton, J.R., Phillips, B.A., Allen, D.A., 1979, MNRAS, 187, 813

\bibitem[]{}
Belczy{\'n}ski, K., Miko{\l}ajewska, J., 1998, MNRAS, 296, 77

\bibitem[]{}
Bromage, G.E., Nandy, K., 1973, A\&A, 26, 17

\bibitem[1916]{can16}
Cannon, A.J., 1916, Ann. Harv. Coll. Obs., 76, 19

\bibitem[]{}
Cant{\'o}, J., Elliott, K.H., Meaburn, J., Theokas, A.C., 1980, MNRAS, 193, 911

\bibitem[1990]{caetal90}
Carter, B.S., 1990, MNRAS, 242,1

\bibitem[1982]{caetal82}
Cassatella, A., Ponz, D., Selvelli, P.L., 1982, NASA IUE
Newslett. No. 10, 31

\bibitem[]{}
Chaffee, F.H. (Jr.), White, R.E., 1982, ApJS, 50, 169

\bibitem[1986]{cletal86}
Clavel, J., Gilmozzi, R., Prieto, A., 1986, NASA IUE
Newslett. No. 31, 83

\bibitem[1969]{cm69}
Cox, D.P, Mathews, W.G., 1969, ApJ, 155, 859

\bibitem[]{}
Dobrzycka, D., Kenyon, S.J., Milone, A.A.E., 1996b, AJ, 111, 414

\bibitem[1996]{detal96}
Dobrzycka, D., Kenyon, S.J., Proga, D., Miko{\l}ajewska, J.,
Wade, R.A., 1996a, AJ, 111, 2090

\bibitem[]{}
Drake, S.A., Ulrich, R.K., 1980, ApJS, 42, 351

\bibitem[1973]{eg73}
Eggen, O.J., 1973, PASP, 85, 42

\bibitem[]{}
Evans, A., Callus, C.M., Albinson, J.S., Whitelock, P.A.,
Glass, I.S., Carter, B., Roberts, G., 1988, MNRAS, 234, 755

\bibitem[]{}
Ezuka, H., Ishida, M., Makino, F., 1998, ApJ, 499, 388

\bibitem[1977]{fetal77}
Feast, M.W., Robertson, B.S., Catchpole, R.M., 1977,
MNRAS, 179, 499

\bibitem[1989]{fgwc}
Feast, M.W., Glass, I.S., Whitelock, P.A., Catchpole, R.M., 1989,
MNRAS, 241, 375

\bibitem[]{}
Fireman, G., Imhoff, C., 1989, IUE NASA Newsletter, 40, 10

\bibitem[1912]{flem12}
Fleming, W.P., 1912, Ann. of HCO, 56, 165

\bibitem[1989]{fpc89}
de Freitas Pacheco, J.A., Costa, R.D.D., 1992,  A\&A, 257, 619 (FPC92)

\bibitem[]{}
Gallagher, J.S., Holm, A.V., Anderson, C.M., Webbink, R.F., 1979,
ApJ, 229, 994

\bibitem[]{}
Girard, T., Willson, L.A., 1987, A\&A, 183, 247

\bibitem[1995]{gl95}
Glass, I.S.,  Whitelock, P.A., Catchpole, R.M., Feast, M.W.,
1995, MNRAS, 273, 383

\bibitem[]{}
Haniff, C.A., Scholtz, M., Tuthill, P.G., 1995, MNRAS, 276, 640

\bibitem[1996]{hh96}
Harries, T.J., Howarth, I.D., 1996, A\&AS, 119, 61

\bibitem[1975]{hl75}
Hayes, D., Latham, D., 1975, ApJ, 197, 593

\bibitem[1976]{hen76}
Henize, K.G., 1976, ApJS, 30, 491

\bibitem[]{}
Herbig, G.H., 1993, ApJ, 407, 142

\bibitem[]{}
Hollis, J.M., Oliversen, R.J., Kafatos, M., Michalitsianos, A.G.,
1986, ApJ, 301, 877

\bibitem[1982]{hoetal82}
Holm, A., Bohlin, R.C., Cassatella, A., Ponz, D., Schiffer, F.H.,
1982, A\&A, 112, 341

\bibitem[]{}
Iben, I., Tutukov, A.V., 1996, ApJS, 105, 145

\bibitem[]{}
Imhoff, C., Wasatonic, R., 1986, IUE NASA Newsletter, 29, 45

\bibitem[1994]{is94}
Ivison, R.J., Seaquist, E.R., 1994, MNRAS, 268, 561 (IS94)

\bibitem[]{}
Josafatsson, K., Snow, T.P., 1987, ApJ, 319, 436.

\bibitem[]{}
Kafatos, M., Lynch, J.P., 1980, ApJS, 42, 611

\bibitem[1982]{kaf82}
Kafatos, M., Michalitsianos, A.G., Feibelman, W.A., 1982, ApJ, 257, 204 (K82)


\bibitem[1985]{kaf85}
Kafatos, M., Michalitsianos, A.G., Fahey, R.P., 1985, ApJS, 59, 785 (K85)

\bibitem[]{}
Karovska, M., Hack, W., Raymond, J., Guinan, E., 1997, ApJ, 482, L175

\bibitem[]{}
Kato, M., 1997, in Physical Processes in Symbiotic Binaries and Related
Systems, ed. J. Miko{\l}ajewska, Copernicus Foundation for Polish Astronomy,
Warsaw, p.65

\bibitem[]{}
Keenan, F.P., Johnson, C.T., Kingston, A.E., 1988, A\&A, 202, 253

\bibitem[]{}
Keenan, F.P., Aller, L.H., Hyung, S., Conlon, E.S., Warren, G.A., 1993,
ApJ, 410, 430

\bibitem[]{}
Keenan, F.P., Aller, L.H., Hyung, S., Brown, P.J.F., 1995, PASP, 107, 148

\bibitem[1986]{kenyon}
Kenyon, S.J. 1986, The Symbiotic Stars, Cambridge Univ. Press,
Cambridge

\bibitem[1988]{kfc88}
Kenyon, S.J., Fernandez--Castro, 1987, AJ, 93, 938

\bibitem[1988]{kfcs88}
Kenyon, S.J., Fernandez--Castro, T., Stencel, R.E., 1988, AJ, 95, 1817

\bibitem[1991]{kommsga91}
Kenyon, S.J., Oliversen, N.A., Miko{\l}ajewska, J., Miko{\l}ajewski, M.,
Stencel, R.E., Garcia, M.R., Anderson, C.M., 1991, AJ, 101, 637

\bibitem[]{}
Kenyon, S.J., Miko{\l}ajewska, J., Miko{\l}ajewski, M., Polidan, R.S.,
Slovak, M.H., 1993, AJ, 106, 1573

\bibitem[1979]{k79}
Klutz, M., 1979, A\&A, 73, 244

\bibitem[1981]{ks81}
Klutz, M., Swings, J.P., 1981, A\&A, 96, 406

\bibitem[1978]{ketal78}
Klutz, M., Simonetto, O., Swings, J.P., 1978, A\&A, 66, 283

\bibitem[]{}
Lim, J., 1993, private communication

\bibitem[]{}
Livio, M., 1988,
in The Symbiotic Phenomenon, eds. J.  Miko{\l}ajewska et al.,
Kluwer, Dordrecht, p.149

\bibitem[]{}
Mattei, J.A., 1997, JAAVSO, 25, 57

\bibitem[]{}
Michalitsianos, A.G., Kafatos, M., Hobbs, R.W., 1980, ApJ, 237, 506

\bibitem[]{}
Michalitsianos, A.G., Kafatos, M., Fahey, R.P., Viotti, R.,
Cassatella, A., Altamore, A., 1988, ApJ, 331, 477

\bibitem[]{}
Michalitsianos, A.G., Kafatos, M., Meier, S.R., 1992, ApJ, 389, 649

\bibitem[]{}
Miko{\l}ajewska, J., 1994, in Interacting Binary Stars, ed. A.W. Shafter,
ASP conf. Series, Vol. 56, p.374

\bibitem[1992]{mk92a}
Miko{\l}ajewska, J., Kenyon, S.J., 1992a, AJ, 103, 579

\bibitem[1992]{mk92b}
Miko{\l}ajewska, J., Kenyon, S.J., 1992b, MNRAS, 256, 177

\bibitem[1987]{metal87}
Miko{\l}ajewska, J., Miko{\l}ajewski, M., Biernikowicz, R.,
Selvelli, P.L., Tur{\l}o, Z., 1987, in Circumstellar Matter, eds.
I. Appenzeller \& C. Jordan, D.Reidel, p.487

\bibitem[1988]{msh88}
Miko{\l}ajewska, J., Selvelli, P.L., Hack, M., 1988, A\&A, 198, 150

\bibitem[1995]{metal95}
Miko{\l}ajewska, J., Kenyon, S.J., Miko{\l}ajewski, M., Garcia, M.R.,
Polidan, R.S., 1995, AJ, 109, 1289

\bibitem[1997]{mas97}
Miko{\l}ajewska, J., Acker, A., Stenholm, B., 1997, A\&A, 327, 191

\bibitem[]{}
M{\"u}ller, B.E., Nussbaumer, H., 1985, A\&A, 145, 144

\bibitem[]{}
Munari, U., Whitelock, P.A., 1989, MNRAS, 237, 45P

\bibitem[]{}
Munari, U., Zwitter, T., 1997, A\&A, 318, 269

\bibitem[1992]{m92}
Munari, U., Yudin, B.F., Taranova, O.G., Massone, G.,
Marang, F., Roberts, G., Winkler, H., Whitelock, P.A., 1992,
A\&AS, 93, 383

\bibitem[1994]{m:n}
M{\"u}rset, U., Nussbaumer, H. 1994, A\&A, 282, 586



\bibitem[1991]{mnsv91}
M{\"u}rset, U., Nussbaumer, H., Schmid, H.M., Vogel, M., 1991,
A\&A, 248, 458

\bibitem[1997]{mwj97}
M{\"u}rset, U., Wolff, B., Jordan, S., 1997, A\&A, 319, 201

\bibitem[]{}
Neckel, Th., Klare, G., 1980, A\&AS, 42, 251

\bibitem[1975]{n75}
Netzer, H., 1975, MNRAS, 171, 395

\bibitem[1996]{njvh96}
Nota, A., Jedrzejewski, R., Voit, M., Hack, W., 1996, FOC Instrument
Handbook V7.0, (Space Telescope Science Institute)

\bibitem[]{}
Nussbaumer, H., 1982, in The Nature of Symbiotic Stars,
eds. M. Friedjung \& R. Viotti, D. Reidel, p.85

\bibitem[]{}
Nussbaumer, H., Storey, P.J., 1978, A\&A, 70, 37

\bibitem[]{}
Nussbaumer, H., Storey, P.J., 1979, A\&A, 71, L5

\bibitem[]{}
Nussbaumer, H., Storey, P.J., 1981, A\&A, 99, 177


\bibitem[]{}
Nussbaumer, H., Storey, P.J., 1982, A\&A, 115, 205


\bibitem[]{}
Nussbaumer, H., Storey, P.J., 1984, A\&AS, 56, 293

\bibitem[]{}
Nussbaumer, H., Vogel, M., 1990, A\&A, 236, 117

\bibitem[]{}
Nyman, L.-${\rm \AA.}$, Olofsson, H., 1986, A\&A, 158, 67

\bibitem[1993]{o93}
Orio, M., 1993, A\&A, 274, L41

\bibitem[]{}
Paczy{\'n}ski, B., 1970, Acta Astron., 20, 47

\bibitem[]{}
Paresce, F., 1990, ApJ, 357, 231

\bibitem[]{}
Patterson, J., Raymond, J.C., 1985, ApJ, 292, 535

\bibitem[]{}
Payne-Gaposchkin, C., Gaposchkin, S., 1938, Variable Stars, (Cambridge:
Harvard University Press)

\bibitem[1897]{pic97}
Pickering, E.C., 1897, Circ. Harv. Coll. Obs., No.17

\bibitem[1914]{pic14}
Pickering, E.C., 1914, Circ. Harv. Coll. Obs., No.182

\bibitem[]{}
Prialnik, D., Kovetz, A., 1995, ApJ, 445, 789

\bibitem[]{}
Proga, D., 1994, private communication

\bibitem[1994]{pmk}
Proga, D., Miko{\l}ajewska, J., Kenyon, S.J. 1994, MNRAS, 268, 213

\bibitem[1996]{pkrm}
Proga, D., Kenyon, S.J., Raymond, J.C., Miko{\l}ajewska, J. 1996,
ApJ, 471, 930

\bibitem[]{}
Reimers, D., Cassatella, A., 1985, ApJ, 297, 275

\bibitem[1973]{ss73}
Sanduleak, N., Stephenson, C.B., 1973, ApJ, 185, 899


\bibitem[1989]{schmid}
Schmid, H.M. 1996, MNRAS, 282, 511

\bibitem[1990]{ss90a}
Schmid, H.M., Schild, H., 1990a, MNRAS, 246, 84

\bibitem[1990]{ss90b}
Schmid, H.M., Schild, H., 1990b, A\&A, 236, L13

\bibitem[1989]{sch89}
Sch{\"o}nberner, D., 1989, in Planetary Nebulae, ed. S. Torres-Peimbert,
Kluwer, p.463

\bibitem[]{}
Schulte-Ladbeck, R.E., 1988, A\&A, 189, 97

\bibitem[]{}
Schwarz, H.E., Nyman, L.-{\AA}., Seaquist, E.R., Ivison, R.I., 1995,
A\&A, 303, 833

\bibitem[1979]{seat}
Seaton, M.J. 1979, MNRAS, 187, 73P

\bibitem[]{}
Seaquist, E.R., Taylor, A.R., 1987, ApJ, 312, 813

\bibitem[]{}
Seaquist, E.R., Taylor, A.R., 1992, ApJ, 387, 624

\bibitem[]{}
Seaquist, E.R., Taylor, A.R., Button, S., 1984, ApJ, 284, 202 (STB84)

\bibitem[]{}
Seaquist, E.R., Krogulec, M., Taylor, A.R., 1993, ApJ, 410, 260

\bibitem[1992]{scg92}
Selvelli, P.L., Cassatella, A., Gilmozzi, R., 1992, ApJ, 393, 289

\bibitem[]{}
Sembach, K.R., Danks, A.C., Savage, B.D., 1993, A\&AS, 100, 107

\bibitem[]{}
Sion, E.M., 1997, in Physical Processes in Symbiotic Binaries and Related
Systems, ed. J. Miko{\l}ajewska, Copernicus Foundation for Polish Astronomy,
Warsaw, p.49

\bibitem[]{}
Smak, J., 1964, ApJS, 9, 141

\bibitem[1977]{s77}
Snow, T.P., York, D.G., Welty, D.E., 1977, AJ, 82, 113

\bibitem[1983]{sb83}
Stone, R.P.S., Baldwin, J.A., 1983, MNRAS, 204, 347

\bibitem[]{}
Swings, J.P., Klutz, M., 1976, A\&A, 46, 303

\bibitem[1941]{ss41}
Swings, P., Struve, O., 1941, ApJ, 94, 291

\bibitem[]{}
Uus, U.H., 1970, Nauch. Inf., 17, 48

\bibitem[]{} 
Van Leeuwen, F., Feast, M.W., Whitelock, P.A., Yudin, B., 1997. MNRAS, 
287, 955

\bibitem[]{}
Van Winckel, H., Duerbeck, H.W., Schwarz, H., 1993, A\&AS, 102, 401

\bibitem[1986]{wal86}
Wallerstein, G., 1986, A\&A, 163, 337

\bibitem[]{}
Webbink, R.F., Livio, M., Truran, J.W., Orio, M., 1987, ApJ, 314, 653

\bibitem[]{}
Webster, B.L., Allen, D.A., 1975, MNRAS, 171, 171

\bibitem[1987]{w87}
Whitelock, P.A. 1987, PASP, 99, 573

\bibitem[1988]{w88}
Whitelock, P.A. 1988,
in The Symbiotic Phenomenon, eds. J.  Miko{\l}ajewska et al.,
Kluwer, Dordrecht, p.47


\bibitem[]{}
Whitelock, P.A., 1998, in Pulsating Stars - Recent Developments in Theory
and Observation, eds. M. Takeuri, D. Sasselov, Universal Academy Press,
Tokyo, astro-ph/9710032

\bibitem[1983]{wh83}
Whitelock, P.A., Catchpole, R.M., Feast, M.W., Roberts, G.,
Carter, B.S., 1983a, MNRAS, 203, 363 (W83)

\bibitem[]{}
Whitelock, P.A., Feast, M.W.,  Catchpole, R.M.,
Carter, B.S., Roberts, G., 1983b, MNRAS, 203, 351

\bibitem[1984]{wh84}
Whitelock, P.A., Menzies, J.W., Lloyd Evans, T., Kilkenny, D.,
1984, MNRAS, 208, 161 (W84)

\bibitem[1994]{wh94}
Whitelock, P.A., Menzies, J., Feast, M., Marang, F., Carter, B.,
Roberts, G., Catchpole, R., Chapman, J., 1994, MNRAS, 267, 711

\bibitem[]{}
Whitelock, P.A., Feast, M.W., Marang, F., Overbeek, M.D., 1997, MNRAS, 288, 512

\bibitem[1924]{yam24}
Yamamoto, I., 1924, Bull. Harv. Coll. Obs., No. 809

\bibitem[]{}
Yudin, B., Munari, U., Taranova, O., Dalmeri, I, 1994, A\&AS, 105, 169

\end{thebibliography}
\end{document}